\documentclass[twocolumn,5p]{autart}
\usepackage{cite}
\usepackage{booktabs}
\usepackage{placeins}
\usepackage{xcolor}
\newif\ifshowrevisions
\showrevisionsfalse
\DeclareRobustCommand{\rev}[1]{{\ifshowrevisions\color{blue}\fi#1}}
\newcommand{\revcolor}{\ifshowrevisions\color{blue}\fi}
\newcommand{\benchmarkresults}{./results_updated}

\usepackage{amsmath,amssymb,amsfonts}
\usepackage{mathtools}
\usepackage{graphicx,subfig}
\newcommand{\precompiledplot}[6]{%
  \begingroup
  \setbox0=\hbox{\includegraphics[pagebox=mediabox]{#1}}%
  \setbox0=\hbox{\kern#5bp\lower\dimexpr#4sp-#6bp\relax\box0}%
  \wd0=#2sp\ht0=#3sp\dp0=#4sp
  \leavevmode\box0
  \endgroup
}

\usepackage{algorithm,algpseudocodex}
\usepackage{textcomp}
\usepackage{pgfplots}
\usepackage{pgfplotstable}
\pgfplotsset{compat=1.18}
\usepgfplotslibrary{groupplots} 
\usepackage{tikz}
\usetikzlibrary{math,arrows.meta,spy,automata,intersections,decorations,arrows,decorations.pathmorphing,decorations.markings,decorations.footprints,fadings,calc,trees,mindmap,shadows,decorations.text,patterns,positioning,shapes,matrix,fit,3d,plotmarks,svg.path}
\usepackage{subcaption}
\usepackage{tikzscale}
\usepackage{arydshln}
\usepackage{pgf-pie}
\pgfdeclarepatternformonly[\GridSize]{my grid}{\pgfqpoint{-1pt}{-1pt}}{\pgfqpoint{\GridSize}{\GridSize}}{\pgfqpoint{\GridSize}{\GridSize}}%
{
     \pgfsetlinewidth{0.25pt}
     \pgfpathmoveto{\pgfqpoint{0pt}{0pt}}
     \pgfpathlineto{\pgfqpoint{\GridSize}{\GridSize + 0.1pt}}
     \pgfpathmoveto{\pgfqpoint{0pt}{0pt}}
     \pgfusepath{stroke}
}

\newdimen\GridSize
\tikzset{
    GridSize/.code={\GridSize=#1},
    GridSize=20pt
}

\tikzset{
    ncbar angle/.initial=90,
    ncbar/.style={
        to path=(\tikztostart)
        -- ($(\tikztostart)!#1!\pgfkeysvalueof{/tikz/ncbar angle}:(\tikztotarget)$)
        -- ($(\tikztotarget)!($(\tikztostart)!#1!\pgfkeysvalueof{/tikz/ncbar angle}:(\tikztotarget)$)!\pgfkeysvalueof{/tikz/ncbar angle}:(\tikztostart)$)
        -- (\tikztotarget)
    },
    ncbar/.default=0.5cm,
}

\tikzset{square left brace/.style={ncbar=0.5cm}}
\tikzset{square right brace/.style={ncbar=-0.5cm}}

\tikzset{round left paren/.style={ncbar=0.5cm,out=120,in=-120}}
\tikzset{round right paren/.style={ncbar=0.5cm,out=60,in=-60}}

\newcommand{\resdir}{./results_updated}
\newcommand{\pdir}{\resdir/paper1}
\newcommand{\Tmain}{8}     % thread level of the main comparison figure
\newcommand{\Nthr}{15}     % horizon of the thread tables/figure

\definecolor{czquick}{HTML}{1F77B4}
\definecolor{cziter}{HTML}{4C78A8}
\definecolor{czplain}{HTML}{9ECAE1}
\definecolor{mptcol}{HTML}{E45756}
\definecolor{yalcol}{HTML}{D4A017}
\definecolor{crcol}{HTML}{4C78A8}
\definecolor{quickcol}{HTML}{76B7B2}
\definecolor{exactcol}{HTML}{59A14F}
\definecolor{gluecol}{HTML}{E15759}
\definecolor{othercol}{HTML}{BAB0AC}

\pgfplotsset{
  only optimizer/.style={
    x filter/.code={%
      \edef\optIdx{\thisrow{optimizerIndex}}%
      \ifnum\optIdx=#1\relax\else\fi}},
  timeaxis/.style={
    width=0.95\linewidth, height=0.52\linewidth,
    xlabel={Prediction horizon \(N\)}, ylabel={Median wall-clock time [s]},
    xmin=2, xmax=15, xtick={2,3,...,15}, grid=both,
    legend columns=3, legend cell align=left,
    legend style={at={(0.5,1.02)},anchor=south,font=\small,draw=none},
    mark size=2pt},
  m5/.style={very thick,mark=o,color=czquick},
  m6/.style={thick,mark=square*,color=cziter},
  m7/.style={thick,mark=triangle*,color=czplain},
  m9/.style={thick,mark=diamond*,color=mptcol},
  m10/.style={thick,mark=pentagon*,color=yalcol},
}
\pgfplotstableset{
  col sep=comma,
  /pgf/number format/1000 sep={\,},
  every head row/.style={before row=\toprule,after row=\midrule},
  every last row/.style={after row=\bottomrule},
}

\newcommand{\summarycurve}[3]{%
  \IfFileExists{\resdir/threads_#1/benchmark_timing_summary.csv}{%
    \addplot+[m#2] table[x=N,y=medianElapsed,col sep=comma,only optimizer=#2]
      {\resdir/threads_#1/benchmark_timing_summary.csv};
    \addlegendentry{#3}}{}}

\newcommand{\pcurve}[5]{%
  \IfFileExists{\pdir/#1}{%
    \addplot+[#4,line width=1.5pt] table[x=#2,y=#3,col sep=comma]{\pdir/#1};
    \addlegendentry{#5}}{}}

\newcommand{\getpct}[2]{%
  \pgfplotstablegetelem{0}{#2_pct}\of{\pietab}\pgfmathsetmacro#1{\pgfplotsretval}}
\newcommand{\mklabel}[2]{% label only for slices of at least 4%
  \pgfmathtruncatemacro\tmpint{round(#2)}%
  \ifdim #2pt<4pt\def#1{}\else\edef#1{\tmpint\%}\fi}
\newcommand{\timepie}[2]{%
  \IfFileExists{\pdir/#1}{%
    \pgfplotstableread[col sep=comma]{\pdir/#1}\pietab
    \getpct\pCR{cr}\getpct\pQ{quick}\getpct\pE{exact}\getpct\pL{loop}%
    \pgfmathsetmacro\pB{max(0,100-\pCR-\pQ-\pE-\pL)}%
    \pgfmathsetmacro\pQ{max(\pQ,0.01)}% pgf-pie: no zero-width slice
    \pgfplotstablegetelem{0}{total}\of{\pietab}\pgfmathsetmacro\pT{\pgfplotsretval}%
    \mklabel\lCR\pCR\mklabel\lQ\pQ\mklabel\lE\pE\mklabel\lL\pL\mklabel\lB\pB
    \begin{tikzpicture}[font=\footnotesize]
      \pie[radius=1.45,sum=auto,hide number,text=inside,
           color={crcol,quickcol,exactcol,gluecol,othercol}]
        {\pCR/\lCR, \pQ/\lQ, \pE/\lE, \pL/\lL, \pB/\lB}
      \node[below] at (0,-1.55) {#2: \pgfmathprintnumber[fixed,precision=1]{\pT}\,s};
    \end{tikzpicture}}{\fbox{\footnotesize missing \texttt{\detokenize{#1}}}}}

\usepackage{enumitem}

\newcommand{\bbm}{\begin{bmatrix}}
\newcommand{\ebm}{\end{bmatrix}}
\newcommand{\bse}{\begin{subequations}}
\newcommand{\ese}{\end{subequations}}
\newcommand{\set}[1]{\mathbb{#1}}
\newcommand{\re}[1]{\mathbb{R}^{#1}}
\newcommand{\Null}{\mathrm{null}}

\newcommand\xqed[1]{%
  \leavevmode\unskip\penalty9999 \hbox{}\nobreak\hfill
  \quad\hbox{#1}}
\newcommand\eor{\xqed{}}

\usepackage{flushend}

\usepackage{enumitem}

\usepackage{hyperref}
\hypersetup{hidelinks=true}

\usepackage{xparse}
\usepackage{marginnote}

\NewDocumentCommand{\hlight}{m O{0pt} m}{%
#3%
}
\protected\def\Scedil{\c{S}}
\protected\def\oumlaut{\"o}
\protected\def\us{\_}

\begin{document}

\begin{frontmatter}

% \title{A symmetry-exploiting approach to\\Explicit MPC for the Constrained Zonotope case\thanksref{footnoteinfo}}
% \title{Accelerating the computation of\\Explicit MPC for the Constrained Zonotope case\thanksref{footnoteinfo}}
\title{Explicit MPC for the constrained zonotope case\\with low-rank matrix updates\thanksref{footnoteinfo}}
\thanks[footnoteinfo]{Corresponding author: Florin Stoican.}

\author[UPB]{\Scedil tefan S. Mihai}\ead{sergiu\us stefan.mihai@upb.ro},
\author[UPB]{Florin Stoican}\ead{florin.stoican@upb.ro},
\author[RUB]{Martin M\oumlaut nnigmann}\ead{martin.moennigmann@rub.de},
\author[UPB]{Bogdan D. Ciubotaru}\ead{bogdan.ciubotaru@upb.ro}
\address[UPB]{Department of Automatic Control and Systems Engineering, \\ Nat. Univ. of Science and Technology Politehnica Bucharest, 060042, Bucharest, Romania.}
\address[RUB]{Automatic Control and System Theory, Ruhr University Bochum, 44801, Bochum, Germany.}


\begin{abstract}
Solving the explicit Model Predictive Control (MPC) problem requires enumerating all critical regions and their associated feedback laws, a task that scales exponentially with the system dimension and the prediction horizon, as well. \hlight{R8-1}{Taking the problem's constraints as polyhedra, the feasible domain is shown to admit a compact constrained-zonotope representation.} Building on this insight, we exploit the geometric properties of the equivalent constrained-zonotope reformulation to accelerate the computation of the explicit solution \hlight{R3-3}{in the linear case}. Specifically, we formulate the multi-parametric problem in the lifted generator space and solve it using second-order optimality conditions, employ low-rank matrix updates to reduce computation time, and introduce an analytic enumeration of candidate active sets that yields the explicit solution in tree form.
\end{abstract}
\end{frontmatter}

%%%%%%%%%%%%%%%%%%%%%%%%%%%%%%%%%%%%%%%%%%%%%%%%%%%%%%%%%%%%%%

\section{Introduction}
\vspace{-2.5mm}

Model Predictive Control (MPC) is a feedback control strategy that employs a dynamic model to predict the evolution of the system states and to compute an optimal sequence of constrained control actions over a finite prediction horizon~\cite{levine_handbook_2018}. \hlight{R1-3}{Significant effort has been devoted to exploiting problem structure~\cite{ferreau2014qpoases} and to developing general-purpose solver improvements~\cite{goulart2026clarabel,lishkova2025successive}. In contrast, another prominent research direction focuses on the \emph{explicit formulation} of MPC, which exploits the multi-parametric structure of the underlying optimization problem~\cite{bemporad2002explicit,alessio2009survey,oberdieck_popparametric_2016,oravec2022real,yang2025lifting}.} \hlight{R3-3}{For the linear (in constraints and dynamics)-quadratic (in cost) MPC problem}, the resulting control law is a piecewise affine function with polyhedral support, partitioning the state space into bounded convex polyhedral regions, the \emph{critical regions}.\vspace{-2.5mm}

The main challenge in explicit MPC is the exponential growth of the critical regions with the prediction horizon. Hence, research has focused on faster implementations, many based on enumerating candidate active constraint sets~\cite{oberdieck_popparametric_2016,ahmadi2018combinatorial,monnigmann2019structure}. Noteworthy, these sets are linked with the face lattice of the polyhedral feasible domain, which can be exploited in the process of solution finding~\cite{seron2003characterisation,mihai2022link}.
% Martin: Had to read the following sentence a couple of times, maybe: Since handling polyhedra quickly becomes intractable for growing system dimension and number of constraints, several alternatives have been considered.
% Stefan: thank you for your suggestion. Indeed, it is much more clear now

As the system dimension and the number of constraints increase, polyhedral computations rapidly become intractable, motivating the search for alternative set representations. \hlight{R8-2}{Owing to their favorable scaling properties and resilience to the curse of dimensionality, constrained zonotopes~\cite{scott2016constrained} have recently been employed in a variety of set-theoretic and control applications, including reachability analysis~\cite{althoff2010computing}, projection algorithms~\cite{vinod2025projection},} \hlight{R8-4}[-1em]{hierarchical MPC~\cite{raghuraman2020hierarchical}, \hlight{R9-1}{tube-based MPC~\cite{andrade2024tube}}, and sparsity-exploiting computational methods~\cite{robbins2025sparsity}}. Furthermore, constrained zonotopes can approximate any convex set arbitrarily well~\cite[Thm.~1]{scott2016constrained} and are closed under fundamental set operations such as affine transformations, Minkowski addition and intersection.
% As the system dimension and number of constraints grow, handling polyhedra quickly becomes intractable, motivating alternative formulations. Thanks to their resilience to the ``curse of dimensionality,'' constrained zonotopes~\cite{scott2016constrained} have been recently used to improve set constructions: reachability analysis~\cite{althoff2010computing}, projection~\cite{vinod2025projection} but also in MPC-related applications such as \hlight{R8-4}{hierarchical MPC \cite{raghuraman2020hierarchical}} or, \hlight{R9-1}{tube MPC implementations \cite{andrade2024tube}, or even to reduce computation time via sparsity exploitation \cite{robbins2025sparsity}}. They can approximate any convex set arbitrarily well~\cite[Theorem~1]{scott2016constrained} and are closed under intersection, Minkowski addition, and affine transformations.\vspace{-2.5mm}

We propose a novel explicit MPC that exploits the geometric properties of the equivalent constrained-zonotope reformulation, extending our previous work~\cite{stoican2024computing}. First, the multi-parametric problem is formulated in the lifted generator space and solved using second-order optimality conditions. Second, computation is accelerated via low-rank matrix updates, thus reducing the time required to compute each critical region. Finally, we introduce an analytic enumeration of candidate active sets that naturally yields the explicit solution in tree form.\vspace{-2.5mm}

The paper is organized as follows. Sec.~\ref{sec:framework} introduces the required definitions. Sec.~\ref{sec:CZEMPC} presents the constrained-zonotope reformulation of the MPC problem. Sec.~\ref{sec:explicit_solution} outlines the computation of the explicit solution and Sec.~\ref{sec:algorithm} describes the associated algorithms. Sec.~\ref{sec:simulation} provides numerical results, with  conclusions in Sec.~\ref{sec:con}.

% We show that our algorithms are comparable with state-of-the-art tools such as YALMIP \cite{lofberg2004yalmip} or MPT3 \cite{herceg2013multi}.

\emph{Notation}
\begin{description}[labelwidth =\widthof{$\Lambda(V)$}, leftmargin=2.75em]
    % \item[$\mathbb{R}^n$] the $n$-dimensional real vector space
    \item[$\set A$] a finite, sorted set of natural numbers
    % \item[$\operatorname{rank}_{{\set A}}(i)$] The rank of $i$ in the set, $\operatorname{rank}_{{\set A}}(i) = \bigl|\{\ell\in{\set A}:\ \ell\le i\}\bigr|$
    \item[$V_{\set A}$] the sub-matrix formed from the rows of matrix $V$ indexed by the elements in $\set A$
    \item[$V_{(\set A)}$] a matrix computed in relation to $\set A$
    \item[$V^+$] the Moore–Penrose generalized inverse of $V$
    \item[$v^{(k)}$] the $k$-th column of the matrix $V$
    \item[$\Lambda(V)$] the spectrum of $V$, i.e., $\{\lambda \in \mathbb{C} : \det(\lambda I - V) = 0\}$
    \item[$e_k$] the $k$-th vector of the canonical basis of $\mathbb{R}^n$
    \item[$n_{\bullet}$] the number of elements in $\bullet$ (e.g., $n_{\set A} = |\set A|$)
    \item[$\otimes$] the Kronecker product
    \item[$x_{i|j}$] the predicted value of variable $x$ at time $i\geq j$, based on information available at time $j$
    \item[$\circ$] the Hadamard product
    \item[$\times$] \hlight{R10-1}[-1.25em]{the Cartesian product} 
\end{description}

\section{Context and framework}\label{sec:framework}
We summarize the standard definitions of polyhedral and constrained zonotopic sets, which will be used later to describe the explicit MPC formulation. Additional details can be found in~\cite{fukuda_polyhedral_2020,scott2016constrained,raghuraman2022set}.

\begin{defn}[Polyhedral set]
\label{def:polyhedron}
The matrix pair $(A\in\mathbb{R}^{d_H\times n},b\in\mathbb{R}^{d_H})$ gives the half-space representation 
\begin{equation}\label{eq:polyhedron_h}
    P(A,b) = \bigl\{x\in\mathbb{R}^n:\: A_i x\leq b_i, \:\forall i\in\{1,\dots,d_H\}\bigr\},
    % P(A,b) = \bigl\{x\in\mathbb{R}^n:\: A_i^\top x\leq b_i, \:i=1:d_H\bigr\}.
\end{equation}
where $b_i$, $A_i$ denote the $i$-th element of vector $b$, and the $i$-th 
row % \textcolor{red}{row} % line
of matrix $A$, respectively. A bounded polyhedral set is called a \textit{polytope}.
\end{defn}

% \textcolor{red}{Def.\ 2 is not really a definition, I think.}
% \begin{defn}[Polytope face] \label{def:face}
A bounded \textit{face} $F$ of the polytope $P$, defined as in Def.~\ref{def:polyhedron}, admits a half-space representation obtained by separating the active and inactive constraints, indexed by the sets $\set A$ and $\set I$, respectively, where $\set A \cup \set I = \{1,\dots,d_H\}$ and $\set A \cap \set I = \emptyset$. The face $F$ is then explicitly given by
\begin{multline}\label{eq:face_halfspace}
    F(\set A) = \bigl\{\, x \in \mathbb{R}^n : A_i x = b_i,\; \forall i \in \set A,\\
    A_j x \leq b_j,\; \forall j \in \set I \,\bigr\}.
\end{multline}
% \end{defn}
% \textcolor{red}{Def.\ 3 also is not really a definition, I am afraid.}
% \begin{defn}[Face lattice]
The collection of all faces forms the face lattice of the polytope, which is a partially ordered set (\textit{poset}). %This lattice admits a graphical representation known as the \textit{Hasse diagram}.
% \end{defn}
% \subsection{Constrained zonotopes}\label{sec:CZ}
% \textcolor{red}{Would probably help to first explain what a zonotope is (informally), then proceed as here with the def.\ of constrained Zs. Or introduce z.\ and constrained z.\ in one def.}
% \textcolor{red}{($\leftarrow$ Would have to be paraphrased for being less of a claim or needs a citation.}

Zonotopes are symmetric polytopes that can be represented as a Minkowski sum of line segments,~\cite[Chap.~4.8]{fukuda_polyhedral_2020}. Although they exhibit strong numerical robustness in high dimensions, they are not closed under set intersection. Constrained zonotopes (CZs) extend zonotopes to overcome this limitation, \cite{scott2016constrained}.
\begin{defn}[Constrained zonotope]
Vectors $c\in\re{n}$, $\theta\in\re{n_c}$, and matrices $G\in\re{n\times n_g}$, $F\in\re{n_c\times n_g}$, define the constrained zonotope
% $\left(c, G, F, \theta\right) \in \re{n}\times\re{n\times n_g}\times\re{n_c\times n_g}\times\re{n_c} $ such that 
\begin{multline}
\label{eq:constr_zonotope}
Z=\left\langle c, G, F, \theta\right\rangle\\
=\left\{x\in \mathbb R^n:\: x=c+G\xi, \:  \|\xi\|_\infty \leq 1, F\xi=\theta \right\}.
\end{multline}
\end{defn}
% Compared to \ref{eq:Grep}, for \textit{CG-rep}, central symmetry is no longer necessary (i.e., any polytope is a constrained zonotope).  

Constrained zonotopes possess several properties~\cite{scott2016constrained}, three of which being of particular relevance in this work. First, they are closed under affine transformations:
% \begin{enumerate}[label=\roman*)]
\begin{equation}
\label{eq:closedMul} 
 r + R Z_1=\left\langle  r+Rc_1, RG_1, F_1, \theta_1\right\rangle.
\end{equation}
Second, CZs are closed under Minkowski sum:
\begin{equation}
\label{eq:closedAdd}
\mkern-25mu  Z_1 \oplus Z_2 = \biggl\langle  c_1+c_2, \bbm G_1 &G_2 \ebm,  \bbm F_1& 0 \\ 0& F_2\ebm, \bbm \theta_1 \\ \theta_2\ebm \biggr\rangle.
\end{equation}
Third, CZs are closed under set intersection:
\vspace{-.5em}
\begin{equation}
\label{eq:closedIntersection}
    \mkern-30mu  Z_1 \cap Z_2 = \left\langle \mkern-4mu c_1, \bbm G_1 & 0 \ebm,\mkern-4mu  \bbm F_1 & \hphantom{-}0 \\ 0 & \hphantom{-}F_2 \\ G_1 & -G_2 \ebm, \bbm \theta_1 \\ \theta_2 \\ c_2 - c_1\ebm \right\rangle,
\end{equation}
% \end{enumerate}
where $Z_1 : = \left\langle  c_1, G_1, F_1, \theta_1\right\rangle$ and $Z_2 : = \left\langle  c_2, G_2, F_2, \theta_2\right\rangle$.

% \begin{rem}
% Repeated applications of set operations~\eqref{eq:closedAdd} and~\eqref{eq:closedIntersection} can result in excessive memory complexity, as they increase the number of generators and constraints. Periodic application of complexity reduction methods is often required in practice~\cite{scott2016constrained,raghuraman2022set}.\eor
% \end{rem}

% Consider a zonotope in $\mathbb R^2$ described by the center $c = \bbm0.15& 0.25\ebm^\top$ and generator matrix $G=\bbm g_1\\ g_2\ebm$, where $g_1 = \bbm-0.75& 0& 1\ebm$ and $g_2 = \bbm 0& 0.5& 0.25 \ebm$. Then, let the plane $\bbm 0.5& -2& 0.25 \ebm \xi = 1$ define the equality constraint; incorporating this constraint as described in \eqref{eq:constr_zonotope} results in the constraint zonotope. Both the original zonotope and the constrained zonotope are depicted in Fig.~\ref{fig:eZono_CZ_Eg}.
\begin{exmp}
% \textcolor{red}{Why not put the example into an example-environment?}
Consider a zonotope in $\mathbb{R}^2$ described by the center
$c=\begin{bsmallmatrix}0.15 & 0.25\end{bsmallmatrix}^{\top}$
and the generator matrix
$G=\begin{bsmallmatrix} g_1 & g_2 & g_3\end{bsmallmatrix}$,
where
$g_1=\begin{bsmallmatrix}-0.75 & 0\end{bsmallmatrix}^\top$, $g_2=\begin{bsmallmatrix}0 & 0.5\end{bsmallmatrix}^\top$,
and
$g_3=\begin{bsmallmatrix}1 & 0.25\end{bsmallmatrix}^\top$.
Then, let the plane
$F\,\xi=\theta$, where $F=\begin{bsmallmatrix}0.5 & -2 & 0.25\end{bsmallmatrix}$ and $\theta=1$,
define the equality constraint; incorporating this constraint as described in
\eqref{eq:constr_zonotope} results in the constrained zonotope. 
% \textcolor{red}{Constrained needs to be related to notation $F\xi= \theta$ in \eqref{eq:constr_zonotope}, I think.}

% \begin{figure}[h]
%     \centering
%     \includegraphics[width=0.4\columnwidth]{./images/zono_cz_example.pdf}
%     \caption{Example of a zonotope and the constrained zonotope}
%     \label{fig:eZono_CZ_Eg}
% \end{figure}

\begin{figure}[h]
  \centering  
    \subfloat[]{\label{fig:cube_plane}\includegraphics[width=.5\columnwidth]{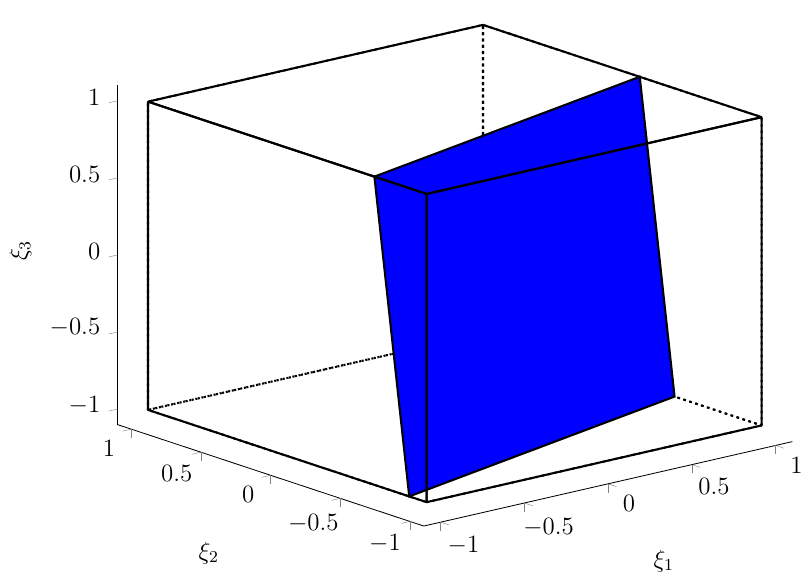}}\hfill
    \subfloat[]{\label{fig:eZono_CZ_Eg}\includegraphics[width=.5\columnwidth]{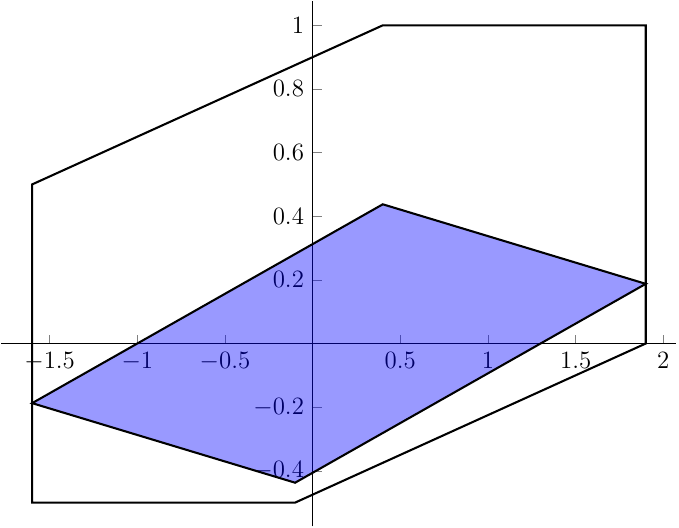}}
    \caption{(a) Domain of $\xi\in\mathbb{R}^{n_g}$, $n_g=3$, and hyperplane defined by $F\xi= \theta$; (b) Zonotope (black contour) and constrained zonotope (blue) in $\mathbb{R}^n$, $n= 2$.
    }
\end{figure}
% \textcolor{red}{How about this caption: ' (a) ' I would then switch the order of the subfigures and drop the subcaptions. Also the axes in what is now (a) need to be $\xi_i$.}
Both the zonotope $\langle c, G, \emptyset, \emptyset \rangle$ and the constrained zonotope $\langle c, G, F, \theta \rangle$ are illustrated in Fig.~\ref{fig:eZono_CZ_Eg}. The key difference lies in the admissible values of $\xi$ used to generate points of the form $c + G\xi$. While the zonotope maps the entire hypercube $\|\xi\|_\infty \leq 1$, the constrained zonotope restricts $\xi$ to the intersection of this hypercube with a hyperplane, as shown in Fig.~\ref{fig:cube_plane}.
\end{exmp}

\section{Constrained zonotope formulation}\label{sec:CZEMPC}

% \subsection{Preliminaries}
% \label{sec:eMPC}

Consider a discrete-time, linear time-invariant system characterized by the state-space representation
    \begin{align}\label{eq:discrete_ss}
        x_{k+1} &= A_d x_k + B_d u_k,
    \end{align}
where $x_k \in \mathbb R^n$ and $u_k \in \mathbb R^m$ denote the state and input vectors at time step $k$, and $A_d\in\mathbb R^{n\times n}$, $B_d\in\mathbb R^{n\times m}$ are the state and input matrices. The state and input constraints are represented as set membership inclusions: $x_k \in \mathcal{X}\subset \mathbb R^n, u_k \in \mathcal{U}\subset \mathbb R^m$, where $\mathcal{X}$ and $\mathcal{U}$ are polytopes.
% \textcolor{red}{'... where $\mathcal{X}$ and $\mathcal{U}$ are polyhedra.' (or maybe polytopes are required?)}
To control such systems, while ensuring constraint satisfaction and cost minimization over a finite prediction horizon of length $N$, a well-established approach is MPC\footnote{To simplify notation, we use the shorthand $x_{i|k} = x_{k+i|k}$ and $u_{i|k} = u_{k+i|k}$ throughout the paper. Whenever clear from context, we denote $x_{0}=x_{0|k}$.}
% Martin: You already introduced 'MPC':
\cite{levine_handbook_2018}, respectively
\begin{subequations}\label{eq:mpc_problem}
\begin{align}
    \label{eq:mpc_problem_cost}\arg\mkern-24mu\min_{\substack{\\ u_{0|k}, \ldots, u_{N-1|k}\\x_{0|k}, \ldots, x_{N|k}}} x_{N|k}^\top& S x_{N|k} + \mkern-8mu\sum_{i=0}^{N-1}\mkern-4mu \ell(x_{i|k}, u_{i|k})\\
    \text{s.t. } x_{i+1|k} &= A_d x_{i|k} + B_d u_{i|k}, \\
    \label{eq:mpc_problem_constraints} x_{i|k} &\in \mathcal{X},\ u_{i|k} \in \mathcal{U},\\
    \label{eq:mpc_problem_constraints_terminal} x_{N|k} &\in \mathcal T.
\end{align}  
\end{subequations}
% \textcolor{red}{I am afraid the problem statement is too incomplete for Automatica: Need to specify which $i$, need to state all opt vars with the $\arg\min$ operator, need to explain double-index notation.}
The stage cost $\ell(x_{i|k}, u_{i|k})$ is defined as 
$\ell(x_{i|k}, u_{i|k}) = x_{i|k}^\top Q x_{i|k} + u_{i|k}^\top R u_{i|k}$, 
where $Q \succeq 0$ and $R \succ 0$ penalize the deviations in state and control input, respectively. The terminal weighting matrix $S = S^\top \succ 0$, together with the polytopic terminal set $\mathcal{T} \subset \mathbb{R}^n$, ensure recursive feasibility and asymptotic stability~\cite{levine_handbook_2018}. To close the loop, at simulation step $k$, the initial predicted state is set to the current state, $x_{0|k} = x_k$, and the control input applied in~\eqref{eq:discrete_ss} is $u_k = u_{0|k}$.

% \color{red}{$S\ge 0$ probably suffices but I do not know a citation, since typically $S$ is obtained from a Riccati eq which provides $S>0$.}\color{black}
% Martin: $R$ is typically assumed to be p.d., not p.s.d., to ensure strict convexity
% Stefan: Thank you. I have updated the text

% \textcolor{red}{the lifted domain 11b is the intersection of three regions, here call them polyhedra, in fact they are zonotopes}
% \textcolor{red}{$x_k \in \mathcal X, k=0:N-1 \Leftrightarrow \bbm u_N\\ x_0\ebm \in \tilde X$}
% \textcolor{red}{$u_k \in \mathcal U, k=0:N-1 \Leftrightarrow \bbm u_N\\ x_0\ebm \in \tilde U$}
% \textcolor{red}{$x_N \in \mathcal T \Leftrightarrow \bbm u_N\\ x_0\ebm \in \tilde T$}
% \textcolor{red}{overall, it means that $\bbm u_N\\ x_0\ebm \in \tilde X \cap \tilde U \cap \tilde T$}

% \clearpage
% \textcolor{red}{2 pages}
% \begin{itemize}
%     \item the MPC problem as an mpQP
%     \item the constrained zonotope case (detailed proofs)
%     \item reduction procedure to remove redundant generators
%     \item problem statement
% \end{itemize}

Reformulating \eqref{eq:mpc_problem} into the equivalent multi-parametric quadratic program (mpQP) by using the substitution $x_{i|k} = A^i x_{0} + \sum_{j=0}^{i-1} A^{i-1-j}B u_{j|k}$ (see~\cite{borrelli2017predictive}) leads to
% Surprising to see $\tilde{Q}$ in the quadratic term, since it is the Hessian:
\begin{subequations}\label{eq:mp-QP}
\begin{align}
    \label{eq:mp-QP_a} \arg\min_{\mathbf u_{[0:N-1]}} &\frac{1}{2} \mathbf u_{[0:N-1]}^\top \tilde Q \mathbf u_{[0:N-1]} + x_0^\top \tilde H \mathbf u_{[0:N-1]}\\
    \label{eq:mp-QP_b}\text{s.t. }& A \mathbf u_{[0:N-1]} \leq E x_0+b,
\end{align}
\end{subequations}
where $\tilde Q\in\mathbb R^{Nm\times Nm}$, $\tilde H\in\mathbb R^{n\times Nm}$, $A\in\mathbb R^ {q\times Nm}$, $b\in\mathbb R^q$, and $E\in\mathbb R^{q\times n}$ are obtained from \eqref{eq:mpc_problem} through standard matrix manipulations. For further use, denote $\mathbf u_{[0:N-1]} = \begin{bsmallmatrix} u_{0|k}^\top & u_{1|k}^\top & \dots & u_{N-1|k}^\top \end{bsmallmatrix}^\top$, $\mathbf x_{[0:N-1]} = \begin{bsmallmatrix} x_{0|k}^\top & x_{1|k}^\top & \dots & x_{N-1|k}^\top \end{bsmallmatrix}^\top$, and note the relations
\begin{subequations}
\label{eq:shorthand}
\begin{align}
    \mathbf x_{[0:N-1]} &=  \tilde A_{[0:N-1]}x_0 + \tilde B_{[0:N-1]}\mathbf u_{[0:N-1]},\\
    x_{N} &=  \tilde A_N x_0 + \tilde B_{N}\mathbf u_{[0:N-1]},
\end{align}
\end{subequations}
where 
% \begin{subequations}
% \label{eq:mpqp_notation}
% \begin{align}
%     \tilde A_{[0:N-1]} &= \bbm I & {A_d^1}^\top & {A_d^2}^\top & \dots & {A_d^{N-1}}^\top \ebm^\top,\\
%     \tilde A_N& =A_d^N,\\
%     \tilde B_{[0:N-1]} &= \bbm 0 & 0 & \dots & 0\\     
%     B_d&0&\dots &0\\ A_d B_d& B_d& \dots &0\\\vdots& \vdots& \ddots&\vdots\\ A_d^{N-2}B_d& A_d^{N-3}B_d&\dots & 0 \ebm,\\
%     \tilde B_{N} &= \bbm A_d^{N-1}B& A_d^{N-2}B_d&\dots & B_d \ebm.
% \end{align}
% \end{subequations}
\begin{subequations}
\label{eq:mpqp_notation}
\begin{align}
    \tilde A_{[0:N-1]} &= \bbm I & {A_d^1}^\top & {A_d^2}^\top & \dots & {A_d^{N-1}}^\top \ebm^\top,\\
    \tilde A_N& =A_d^N,\\
    % \left[\tilde B_{[0:N-1]}\right]_{i,j}&=\begin{cases}
    %     A_d^{i-j-1}B_d,&i>j\\
    %     0,&i\leq j
    % \end{cases}\forall i,j=0,\dots,N-1\\
    \label{eq:mpqp_notation_c}\tilde B_{[0:N-1]} &= \sum_{i=2}^N\sum_{j=1}^{i-1}(e_i e_j^\top)\otimes (A_d^{i-j-1}B_d),\\
    \tilde B_{N} &= \bbm A_d^{N-1}B_d& A_d^{N-2}B_d&\dots & B_d \ebm.
\end{align}
\end{subequations}

% \clearpage

% and second page
% Noteworthy, in \eqref{eq:mpqp_notation_c}, $e_k\in\mathbb R^N$ is the $k$-th standard basis vector and we assume $A_d^0 = I$.

% In our previous work \cite{mihai2022link,mihai2023computing}, we introduced the concept of \textit{lifted feasible domain}, which enables a reformulation of the original constrained optimization problem. Thus, by transforming \eqref{eq:mp-QP_b} into the equivalent representation
% \begin{multline}
% \label{eq:ldom}
%    \mkern-24mu P\left(\bbm A&-E\ebm,b\right) = \biggl\{v\in\mathbb R^{Nm+n}:\: \bbm A & -E\ebm v \leq b \biggr\},
% \end{multline}
% where $v^\top=\bbm \mathbf u_{[0:N-1]}^\top&x_0^\top\ebm$, we embed the the original problem \eqref{eq:mp-QP} into a higher-dimensional space $\mathbb R^{Nm+n}$, rather than the original $\mathbb R^{Nm}$, effectively eliminating explicit parameter dependencies on $x_0$. 
By revisiting the state and input constraints in \eqref{eq:mpc_problem_constraints} and the terminal constraint \eqref{eq:mpc_problem_constraints_terminal}, and incorporating the shorthand notation from \eqref{eq:shorthand}, the feasible domain of \eqref{eq:mpc_problem} is rewritten as $\tilde{\mathcal X}(x_0) \cap \tilde{\mathcal U} \cap \tilde{\mathcal T}(x_0)$, with
\hlight{\parbox{1cm}{R3-4\\R8-5\\R10-1}}{
\begin{align}
    \label{eq:tilde_sets} \tilde {\mathcal X}(x_0) =&\bigl\{\mathbf u_{[0:N-1]}:\tilde A_{[0:N-1]}x_0+ \\
    \nonumber &\qquad\qquad \tilde B_{[0:N-1]}\mathbf u_{[0:N-1]}\in \mathcal X\times\ldots\times \mathcal X\bigr\},\\
    \nonumber\tilde {\mathcal U} =&\bigl\{\mathbf u_{[0:N-1]}:\:u_{j|k}\in \mathcal U, j\in\{0,\ldots, N-1\}\bigr\},\\
    \nonumber \tilde {\mathcal T}(x_0) = &\left\{\mathbf u_{[0:N-1]}:\:\tilde A_Nx_0+\tilde B_{[0:N-1]}\mathbf u_{[0:N-1]}\in \mathcal T\right\}.
\end{align}
}
% A closer look at \eqref{eq:ldom} reveals that the lifted vector $v$ should inherit constraints from its components $u_N$ and $x_0$, defined in . First, we know that $x_{k+i|k}\in \mathcal X,\ i=0,\dots,N-1$, which implies that $v\in\tilde{\mathcal X}$. When the initial state is actually $x_{k+N|k}\in \mathcal X_N$, then we know $v\in\tilde{\mathcal X}_N$. Lastly, $u_{k+i|k}\in \mathcal U$ imposes  $v\in\tilde{\mathcal U}$. The sets $\tilde{\mathcal X}$, $\tilde{\mathcal X_N}$, and $\tilde{\mathcal U}$ are convex polytopes --and possibly zonotopoes-- derived from the sets ${\mathcal X}$, ${\mathcal X_N}$, and ${\mathcal U}$, respectively, with appropriate dimensions for $v$. Overall, the satisfaction of all these constraints gives
For later use, we denote the equivalent polyhedral representation \hlight{R10-1}{$\tilde {\mathcal X}(x_0) \cap \tilde {\mathcal U}\cap \tilde {\mathcal T}(x_0) = \{\mathbf u_{[0:N-1]}\in \mathbb R^{Nm}:\: A\,\mathbf u_{[0:N-1]}\leq b + Ex_0\}$}, with $A\in \mathbb R^{q \times Nm}$, $b\in \mathbb R^{q}$ and $F\in \mathbb R^{q\times n}$, appropriately constructed.

Next, we provide a constrained zonotope description of the feasible domain associated with \eqref{eq:mpc_problem}, which is then used to efficiently derive the explicit MPC solution. 
% \begin{rem}
% Equation~\eqref{eq:ldom} represents a combination of affine transformations, summation, and intersection operations, as outlined in \eqref{eq:tilde_sets}. According to the closure properties established in \eqref{eq:closedMul}--\eqref{eq:closedIntersection}, these operations are preserved within the constrained zonotope class. This observation provides the base for the following subsection: starting from initial constrained zonotope representation of the constraint sets $\mathcal X$, $\mathcal U$ and $\mathcal T$, we derive a constrained zonotope description of the lifted feasible domain associated with \eqref{eq:mpc_problem}, which is then used to derive the explicit MPC solution. 
% \end{rem}
% \subsection{The constrained zonotope formulation}
Consider the case where the state and input constraints are represented in zonotopic form\footnote{\label{foot:hyper-rectangles}\hlight{\parbox{1cm}{R1-4, R8-6}}{Our results apply to the general constrained-zonotopic setting. Nevertheless, the state and input constraints are often specified as hyper-rectangles, which admit a direct zonotopic representation. More generally, any polyhedral set can also be represented as a constrained zonotope~\cite{scott2016constrained}.}} as
    $\mathcal X = \langle c_{\mathcal X}, G_{\mathcal X}\rangle, \quad \mathcal U = \langle c_{\mathcal U}, G_{\mathcal U}\rangle$.
% For $\mathcal T$ I would cite 
%@ARTICLE{83532,
%  author={Gilbert, E.G. and Tan, K.T.},
%  journal={IEEE Transactions on Automatic Control}, 
%  title={Linear systems with state and control constraints: the theory and application of maximal output admissible sets}, 
%  year={1991},
%  volume={36},
%  number={9},
%  pages={1008-1020},
%  keywords={Linear systems;Control systems;Constraint theory;Terminology;Tin;Error correction;Stability},
%  doi={10.1109/9.83532}}
Typically, the terminal set $\mathcal T$ is defined as the result of a set recurrence $\Omega_{k+1}=(A_d+B_dK)^{-1}\Omega_k \cap \Omega_0$, \hlight{\parbox{1cm}{R1-5, R3-5}}{where $\Omega_0 =\mathcal X\cap \{x:\: Kx\in \mathcal U\}$, for a predefined stabilizing static gain $K$.} Since this terminal set can be conveniently represented as a constrained zonotope \cite{gheorghe2024computing}, we express it as
    $\mathcal T =\langle c_{\mathcal T}, G_{\mathcal T}, F_{\mathcal T}, \theta_{\mathcal T}\rangle$. For reference, the generator matrices $G_{\{\mathcal X,\mathcal U,\mathcal T\}}$ belong to $\mathbb R^{\{n,m,n\}\times \{g_X, g_U, g_T\}}$, while the constraint parameters $(F_{\mathcal T},\theta_{\mathcal T})$ are elements of $\mathbb R^{c_{\mathcal T}\times g_{\mathcal T}}\times \mathbb R^{c_{\mathcal T}}$. This allows us to express \eqref{eq:tilde_sets} as
\bse
\label{eq:czon_constraints}
\begin{align}
    \label{eq:czon_constraints_a}\mkern-24mu\tilde A_{[0:N-1]}x_0\mkern-4mu+\mkern-4mu \tilde B_{[0:N-1]}\mathbf u_{[0:N-1]}&\mkern-4mu\in\mkern-4mu\langle \mathbf 1_{N}\mkern-4mu\otimes \mkern-4mu c_{\mathcal X}, I_N \mkern-4mu\otimes\mkern-4mu G_{\mathcal X}\rangle,\\
    \label{eq:czon_constraints_b}\mkern-24mu\mathbf u_{[0:N-1]} &\mkern-4mu\in\mkern-4mu\langle \mathbf 1_{N}\mkern-4mu\otimes \mkern-4muc_{\mathcal U}, I_N \mkern-4mu\otimes \mkern-4mu G_{\mathcal U}\rangle,\\
    \label{eq:czon_constraints_c}\mkern-24mu\tilde A_Nx_0+\tilde B_{[0:N-1]}\mathbf u_{[0:N-1]}&\in \langle c_{\mathcal T}, G_{\mathcal T}, F_{\mathcal T}, \theta_{\mathcal T}\rangle.
\end{align}
\ese
Further manipulations based on \eqref{eq:closedMul}--\eqref{eq:closedIntersection} lead to the constrained zonotope description detailed in the next result.

\begin{lem}\label{lem:lemma-xi-subst}
Introducing \hlight{R3-6}{$\xi\in \mathbb R^{N(g_{\mathcal X}+g_{\mathcal U})+g_{\mathcal T}}$} and applying substitution $\mathbf u_{[0:N-1]}=c_{\mathcal D}+G_{\mathcal D}\xi$ in \eqref{eq:mp-QP} leads to the equivalent formulation
\begin{subequations}\label{eq:mp-QP-CZ}
\begin{align}
    \nonumber\arg\min_{\xi} \quad&\frac{1}{2} \xi^\top G_{\mathcal D}^\top \tilde Q G_{\mathcal D}\xi + c_{\mathcal D}^\top\tilde Q G_{\mathcal D}\xi + x_0^\top \tilde H G_{\mathcal D} \xi\\ 
    \label{eq:mp-QP-CZ_a}&\qquad\qquad\qquad\:+ x_0^\top \tilde H c_{\mathcal D} + \frac{1}{2} c_{\mathcal D}^\top \tilde Q c_{\mathcal D}\\  
    \label{eq:mp-QP-CZ_b}\text{s.t. }\quad& F_{\mathcal D}\xi - \theta_{1,\mathcal D} - \theta_{2,\mathcal D}x_0 = \mathbf 0_{\bar D},\\
    \label{eq:mp-QP-CZ_c}& Y \xi - \mathbf{1}_{2\bar D} \leq \mathbf{0}_{2\bar D},
\end{align}
\end{subequations}
where $Y = \bigl[\begin{matrix} I_{\bar D}& -I_{\bar D} \end{matrix}\bigr]^\top$ and 
\begin{align}
    \label{eq:notation_cz}\mkern-20mu c_{\mathcal D} &= \mathbf 1_{N}\otimes c_{\mathcal U},\\
    \nonumber\mkern-20mu G_{\mathcal D} &= \bbm I_N\otimes G_{\mathcal U}& \mathbf 0_{Nn \times (Ng_X + g_T)}\ebm,\\
    \nonumber\mkern-20mu F_{\mathcal D} &= \bbm \tilde B_{[0:N-1]}(I_N \otimes G_{\mathcal U}) & -I_N \otimes G_{\mathcal X} & \mathbf 0_{Nn \times g_T}\\ \tilde B_{N}(I_N \otimes G_{\mathcal U})& \mathbf 0_{n\times Ng_{\mathcal X}} & -G_{\mathcal T}\\ \mathbf 0_{c_{\mathcal T}\times Ng_{\mathcal U}} & \mathbf 0_{c_{\mathcal T}\times Ng_{\mathcal X}} & F_{\mathcal T}\ebm,\\
    \nonumber\mkern-20mu \theta_{1,\mathcal D} &=\mkern-8mu \bbm \mathbf 1_{N}\otimes c_{\mathcal X}\\c_{\mathcal T}\\\theta_{\mathcal T}\ebm\mkern-8mu -\mkern-8mu\bbm \tilde B_{[0:N]}\\ \mathbf 0_{c_{\mathcal T}}\ebm\mkern-4mu(\mathbf 1_{N}\otimes c_{\mathcal U}), \theta_{2,\mathcal D}=\bbm \tilde A_{[0:N]}\\ \mathbf 0_{c_{\mathcal T}\times n}\ebm.
\end{align}
The term $\bar D=N(g_{\mathcal X}+g_{\mathcal U})+g_{\mathcal T}$ denotes the number of generators and $\bar n_c = (N+1)n + c_{\mathcal T}$ the number of equalities; $\tilde A_{[0:N]}$ and $\tilde B_{[0:N]}$ are obtained by vertically concatenating $(\tilde A_{[0:N-1]}, \tilde A_N)$ and $(\tilde B_{[0:N-1]}, \tilde B_N)$, respectively.
\end{lem}
\begin{pf}
Substituting \eqref{eq:czon_constraints_b} into \eqref{eq:czon_constraints_a}, following the approach in \eqref{eq:closedIntersection}, gives the constrained zonotope bounding the control inputs sequence $\mathbf u_{[0:N-1]}$ as
\vspace{-.5em}
        \begin{multline*}
        \biggl\langle \mathbf 1_{N}\otimes c_{\mathcal U}, \bbm I_N\otimes G_{\mathcal U}& \mathbf 0_{Nm \times Ng_X}\ebm, \\ 
        \bbm \tilde B_{[0:N-1]}(I_N \otimes G_{\mathcal U}) & -I_N \otimes G_{\mathcal X}\ebm,\\
        \mathbf 1_{N}\otimes c_{\mathcal X}-\tilde B_{[0:N-1]}(\mathbf 1_{N}\otimes c_{\mathcal U})-\tilde A_{[0:N-1]}x_0\biggr\rangle.
    \end{multline*}    
    Including \eqref{eq:czon_constraints_c} in the previous set and performing a minor regrouping of the terms to highlight $x_0$ yields 
    \begin{multline}
        \label{eq:domain}
        \bigl\{\mathbf u_{[0:N-1]}=c_{\mathcal D}+G_{\mathcal D}\xi,\\
        \: F_D\xi = \theta_{1,\mathcal D}+\theta_{2,\mathcal D}x_0,\: \|\xi\|_\infty\leq 1\bigr\},
    \end{multline}
with $c_{\mathcal D}, G_{\mathcal D}, F_{\mathcal D}$, $\theta_{1,\mathcal D}$ and $\theta_{2,\mathcal D}$ defined as in \eqref{eq:notation_cz}. 

Making the change of variable $\mathbf u_{[0:N-1]}=c_{\mathcal D}+G_{\mathcal D}\xi$ maps \eqref{eq:mp-QP} into \eqref{eq:mp-QP-CZ}, thus concluding the proof.  \hfill\hfill\qed
\end{pf}

\begin{exmp}
\label{ex:cz_versus_polyhedral}
\hlight{\parbox{1cm}{R1-3 R3-7}}{
For the scalar system $x_{k+1}=x_{k}-u_{k}$, with constraints $\mathcal X=\{\,0.8\leq x\leq 1.2\,\}$, $\mathcal U=\{\,|u|\leq 0.3\,\}$ and prediction horizon $N=1$, the feasible domain associated with the singleton input sequence $\mathbf u_{[0:0]}=\{u_0\}$ is obtained from the conditions $u_0\in\mathcal U$ and $x_1\in\mathcal X$. These translate into the inequalities $-0.3\leq u_0\leq 0.3$ and $0.8\leq x_0-u_0\leq 1.2$, respectively. The resulting feasible domain in the $(u_0,x_0)$ space is illustrated in Fig.~\ref{fig:p_illustration}.
Introducing auxiliary variables $\xi_1,\xi_2\in[-1,1]$, the constraints are rewritten as $u_0=0.3\: \xi_1$ (with the substitution in Lem.~\ref{lem:lemma-xi-subst}) and $x_0-u_0=1-0.2\:\xi_2$, which leads to the constrained zonotope representation
\vspace{-1em}
\begin{multline*}
    u_0\in \langle c_D, G_D, F_D, \theta_{1,D}+\theta_{2,D}x_0\rangle\\
    =\bigl\langle 0, [0.3\: 0], [0.3\: -0.2], -1+x_0\bigr\rangle.
\end{multline*}}
\vspace{-3em}
\begin{figure}[!ht]
    \centering

    \subfloat[polyhedral interpretation]{%
        \label{fig:p_illustration}
        \includegraphics[
            width=0.45\columnwidth,
            trim=4.2cm 17.75cm 10.75cm 4.25cm,
            clip
        ]{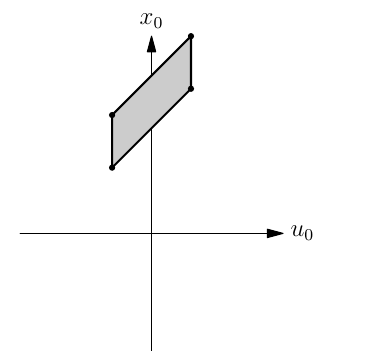}
        \tikz[remember picture,overlay]\coordinate (polyfig) at (0,0);
    }
    \hfill
    \subfloat[CZ interpretation]{%
        \label{fig:cz_illustration}
        \includegraphics[
            width=0.45\columnwidth,
            trim=4.7cm 13.75cm 7.75cm 4.75cm,
            clip
        ]{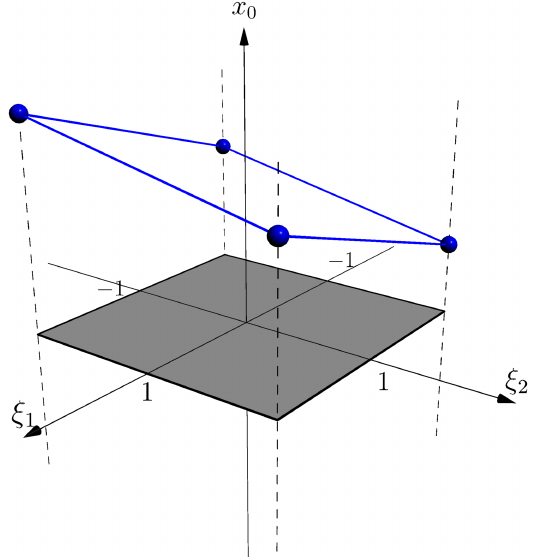}
        \tikz[remember picture,overlay]\coordinate (czfig) at (0,0);
    }

    \begin{tikzpicture}[remember picture,overlay]
        \draw[
            {Latex[length=3mm]-},
            thick,
            bend left=25
        ]
        ($(polyfig)+(-2cm,2cm)$)
        to node[midway, above=0mm] {\small $u_0=c_D+G_D\xi$}
        ($(czfig)+(-3.25cm,2cm)$);
    \end{tikzpicture}

    \caption{The constrained zonotope (CZ) explicit MPC idea.}
    \label{fig:cz_illustration_all}
\end{figure}
\vspace{-1em}

\hlight{}{The geometric interpretation of this inclusion is illustrated in Fig.~\ref{fig:cz_illustration}. Expressing the feasible domain in the space of $(\xi_1, \xi_2, x_0)$ yields a hypercube constraint in the $(\xi_1,\xi_2)$ plane, which is then lifted along the $x_0$ dimension and intersected with an affine subspace defined by the equality constraints coupling all three variables.
 }
\end{exmp}

% \newpage
\section{Solution computation and improvements}\label{sec:explicit_solution}

As per \eqref{eq:domain}, the decision variable $\xi$ lies in the intersection between a fixed hypercube and an affine subspace, parameterized in $x_0$. Hence, we focus on the structural simplicity of \eqref{eq:mp-QP-CZ}, using its constrained zonotope representation, to compute efficiently the explicit solution.

\subsection{The critical region and the associated affine law}

Although the reformulated problem \eqref{eq:mp-QP-CZ} features a quadratic cost with weighting matrix\footnote{\label{foot:qtilde}\hlight{R1-6}{$\tilde Q$ is obtained as a linear combination of matrices $Q, S \succeq 0$, $R \succ 0$. The strict positive definiteness of $R$ ensures that $\tilde Q$ is itself positive definite.}} $\tilde Q\succ 0$, the Hessian $G_{\mathcal D}^\top \tilde Q G_{\mathcal D}$ is not strictly positive definite. Since $G_{\mathcal D} \in \mathbb R^{Nm \times \bar D}$, its rank is at most $Nm < \bar D$, implying rank deficiency. Thus, the Karush--Kuhn--Tucker (KKT) conditions remain necessary but are no longer sufficient. To recover the optimality, second-order conditions must be included to account for the null space of $G_{\mathcal D}^\top \tilde Q G_{\mathcal D}$.

% Recall the notions of null space of a matrix $X\in \mathbb R^{p \times r}$ of rank $p$, with $p<r$, as $\ker(X) = \{z\in \mathbb R^r:\: Xz=0\}$. Let $\{z_1,z_2,\dots,z_k\}$ be a basis of $\ker(X)$, and define $\text{null}(X)$ as a matrix whose columns are vectors from the basis
% \begin{equation}
%     \text{null}(X) = \begin{bmatrix}
%         z_1&z_2&\dots & z_k
%     \end{bmatrix}
% \end{equation}
\vspace{-.5em}
\begin{prop}\label{prop:kkt_z}
For a candidate set of active inequalities $\set A\subset \{1,\ldots, 2\bar D\}$, the optimal solution minimizing \eqref{eq:mp-QP-CZ} leads to the affine law 
    \begin{multline}
        \label{eq:affine_law}\mathbf u_{(\mathbb A), [0:N-1]}^\star(x_0) = G_{\mathcal D}K_{(\set A)}^{-1}\kappa_{2,(\set A)}x_0\\
        +\left(c_{\mathcal D}+G_{\mathcal D}K_{(\set A)}^{-1}\kappa_{1,(\set A)}\right)
    \end{multline}
and its associated (possibly empty) critical region
\begin{multline}
        \label{eq:cr}\text{CR}_{(\set A)}=\Biggl \{\bbm Y_{\set I}K_{(\set A)}^{-1}\kappa_{2,(\set A)}\\ -S_{\set A, 2}\ebm x_0 \\
        + \bbm Y_{\set I}K_{(\set A)}^{-1}\kappa_{1,(\set A)}\\-s_{\set A, 2}\ebm \leq \bbm \mathbf 1_{2\bar D - n_{\set A}}\\\mathbf 0_{n_{\set A\hphantom{D-}}}\ebm\Biggr \},      
    \end{multline}
over which it is active, with the notation
\bse
\label{eq:notation_kkt}
\begin{align}
\label{eq:notation_kkt_KA}\mkern-32mu K_{(\set A)}&=\bbm Z_{(\set A)}^\top G_{\mathcal D}^\top \tilde Q G_{\mathcal D} \\ F_{\mathcal D} \\ Y_{\set A}\ebm,\\
\label{eq:notation_kkt_k12}\mkern-32mu\kappa_{1,(\set A)}&\mkern-4mu=\mkern-4mu \bbm -Z_{(\set A)}^\top G_{\mathcal D}^\top \tilde Q c_{\mathcal D}\\ \theta_{1,\mathcal D}\\ \mathbf{1}_{n_{\set A}} \ebm\mkern-8mu, \kappa_{2,(\set A)}\mkern-4mu=\mkern-4mu\bbm -Z_{(\set A)}^\top G_{\mathcal D}^\top \tilde H^\top\\\theta_{2,\mathcal D}\\ \mathbf{0}_{n_{\set A}} \ebm,\\
\label{eq:notation_kkt_SA}\mkern-32mu S_{(\set A)}&=-T_{(\mathbb A)}^{+} G_{\mathcal D}^\top \bigl(\tilde Q G_{\mathcal D}K_{(\set A)}^{-1}\kappa_{2,(\set A)}+\tilde H^\top\bigr),\\
\label{eq:notation_kkt_sA}\mkern-32mu s_{(\set A)}&=-T_{(\mathbb A)}^{+} G_{\mathcal D}^\top \tilde QG_{\mathcal D}K_{(\set A)}^{-1}\kappa_{1,(\set A)}c_{\mathcal D}, \\
\label{eq:notation_kkt_TZ}\mkern-32mu T_{(\mathbb A)} &= \bbm F_{\mathcal D}^\top & Y_{\set A}^\top\ebm,\:Z_{(\mathbb A)} = \Null\left(T_{(\set A)}^\top \right),
\end{align}
\ese
where $\set I =\{1,\ldots,2\bar D \}\setminus \set A$, $n_{\mathbb A}$ denotes the cardinality of index set $\mathbb A$, and $S_{\set A, 2}$, $s_{\set A, 2}$ are the sub-matrices gathering the last $n_{\set A}$ rows from $S_{(\set A)}$, $s_{(\set A)}$, corresponding to $\mu_{\set A}$.
\end{prop}
\vspace{-2em}
\begin{pf}
\hlight{R10-3}{}As per \cite[Prop. 1.30]{bertsekas2014constrained}, $\xi^\star$ is a strict local minimum of \eqref{eq:mp-QP-CZ} for the given set of active indices $\set A$ \hlight{}{if and only if} there exist the Lagrangian multipliers $\lambda^\star,\mu^\star$ s.t.
\begin{subequations}
\label{eq:empc_kkt_II}
\begin{align}
    \nonumber G_{\mathcal D}^\top \tilde Q G_{\mathcal D}\xi^\star + G_{\mathcal D}^\top \tilde Q c_{\mathcal D} +  G_{\mathcal D}^\top \tilde H^\top x_0\\ + {F_{\mathcal D}}^\top \lambda^\star + Y^\top_{\set A} \mu_{\set A}^\star   &= \mathbf 0_{\bar D},\label{eq:empc_kkt_II_a}\\
        \label{eq:empc_kkt_II_b} F_{\mathcal D}\xi^\star - \theta_{1,\mathcal D} - \theta_{2,\mathcal D}x_0 &=\mathbf 0_{\bar n_c},\\
        \label{eq:empc_kkt_II_c} Y \xi^\star - \mathbf{1}_{2\bar D} &\leq \mathbf 0_{2\bar D},\\
        \label{eq:empc_kkt_II_d}\mu_{\set A}^\star > \mathbf 0_{n_{\mathbb A}},\:\mu_{\set I}^\star &= \mathbf 0_{2\bar D - n_{\mathbb A}},\\
        \label{eq:empc_kkt_II_f}\mu^\star \circ \left(Y \xi^\star - \mathbf{1}_{2\bar D}\right) &= \mathbf 0_{2\bar D},
\end{align}
\end{subequations}
and, for every $z\neq 0$ which satisfies $F_{\mathcal D}z=0$, $Y_{\set A}z=0$, we have that 
\begin{equation}
\label{eq:empc_kkt_II2}
z^\top G_{\mathcal D}^\top \tilde Q G_{\mathcal D} z>0.
\end{equation}
Since, by construction, $\tilde Q\succ 0$ \hlight{R1-6}{(see footnote~\ref{foot:qtilde})}, a necessary and sufficient condition for \eqref{eq:empc_kkt_II2} to hold is that $z$ is in the null spaces of $F_{\mathcal D}$ and $Y_{\set A}$ but not in the null space of $G_{\mathcal D}$, respectively
\begin{equation}
\label{eq:z_subspace}
    z\in \ker F_{\mathcal D} \cap \ker Y_{\set A}, z \notin \ker G_{\mathcal D}.
\end{equation}
The columns of matrix $Z_{(\mathbb A)}$ from \eqref{eq:notation_kkt_TZ}  describe a basis of the subspace \eqref{eq:z_subspace}. Left-multiplying with $Z_{(\mathbb A)}^\top$ in \eqref{eq:empc_kkt_II_a} and reordering s.t. the equalities appear first leads to
\begin{subequations}
\label{eq:kkt_hessian}
\begin{align}
    \label{eq:kkt_hessian_a}\mkern-56mu Z_{(\set A)}^\top\mkern-8mu\left(G_{\mathcal D}^\top \tilde Q G_{\mathcal D}\xi^\star + G_{\mathcal D}^\top \tilde Q c_{\mathcal D} +  G_{\mathcal D}^\top \tilde H^\top x_0\mkern-4mu\right) &= \mathbf 0_{\bar D - \bar n_c - n_{\mathbb A}},\\
        \label{eq:kkt_hessian_b} F_{\mathcal D}\xi^\star - \theta_{1,\mathcal D} - \theta_{2,\mathcal D}x_0 &=\mathbf 0_{\bar n_c},\\
        \label{eq:kkt_hessian_c}Y_{\set A} \xi^\star - \mathbf{1}_{n_{\set A}}&= \mathbf 0_{n_{\set A}},\\
        \label{eq:kkt_hessian_d}\mu_{\set I}^\star &= \mathbf 0_{2\bar D-n_{\set A}},\\
        \label{eq:kkt_hessian_e} Y_{\set I} \xi^\star - \mathbf{1}_{2\bar D-n_{\set A}} &\leq \mathbf 0_{\bar D-n_{\set A}},\\
        \label{eq:kkt_hessian_f}\mu_{\set A}^\star &> \mathbf 0_{n_{\set A}}.
\end{align}
\end{subequations}
Rearranging\footnote{The invertibility of $K_{(\set A)}\in \mathbb R^{\bar D \times \bar D}$ is assumed to hold and it is one of the test conditions appearing later in Algorithm~\ref{alg:mpQP_CZ}.}\eqref{eq:kkt_hessian_a}--\eqref{eq:kkt_hessian_c} to isolate $\xi^\star$ and substituting it into \eqref{eq:empc_kkt_II_b}, yields, using the notation of \eqref{eq:notation_kkt}, both the primal and dual optimal solutions
\begin{equation}\label{eq:KKTsolution}
K_{(\set A)}\xi^\star = \kappa_{2,(\set A)}x_0+\kappa_{1,(\set A)},\ 
\bbm \lambda^\star\\ \mu_{\set A}^\star\ebm = S_{(\set A)}x_0 + s_{(\set A)}.
\end{equation}
Mapping $\mathbf u_{[0:N-1]}^\star(x_0) = c_{\mathcal D}+G_{\mathcal D}\xi^\star(x_0)$ yields the optimal affine law \eqref{eq:affine_law}, while substituting \eqref{eq:KKTsolution} into \eqref{eq:kkt_hessian_e}--\eqref{eq:kkt_hessian_f} provides the description of the critical region, leading to \eqref{eq:cr} and concluding the proof. \hfill\hfill\qed
\end{pf}

When $\Null\!\left(T_{(\set A)}\right) = \{\mathbf 0_{\bar D}\}$, the matrix $Z_{(\set A)}\in \mathbb R^{\bar D\times (\bar D - \bar n_c - n_{\set A})}$ is empty and Prop.~\ref{prop:kkt_z} reduces to the following result.  

\begin{cor}
Let $\set A$ be a candidate active set such that $\bar D = \bar n_c + n_{\set A}$. Then, the KKT conditions are necessary and sufficient to recover from \eqref{eq:mp-QP-CZ} the affine law \eqref{eq:affine_law} and the associated critical region \eqref{eq:cr}, provided that the matrices $K_{(\set A)}$, $\kappa_{1,(\set A)}$, and $\kappa_{2,(\set A)}$ (cf.~\eqref{eq:notation_kkt_KA}--\eqref{eq:notation_kkt_k12}) are redefined by retaining only their last $\bar n_c + n_{\set A}$ rows.
\end{cor}
\begin{figure*}[!ht]    
    % \begin{tikzpicture}[transform canvas={scale=.9}]
    %     \input{./images/iterative_updates.tikz}
    % \end{tikzpicture}
    \includegraphics[width=.9\textwidth]{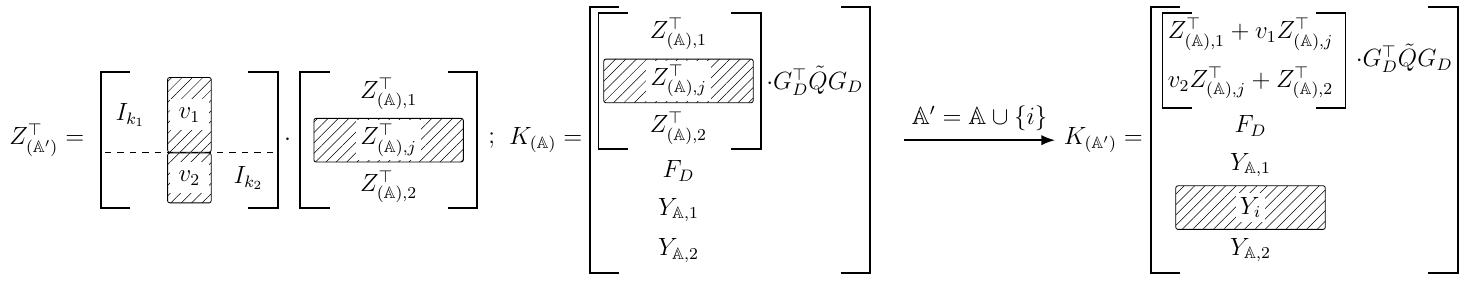}
    \caption{Illustration of the update of matrices $Z_{(\mathbb A)}$, $K_{(\mathbb A)}$.}
    \label{fig:kkt_update}
\end{figure*}  
\begin{pf}
Applying the KKT conditions to \eqref{eq:mp-QP-CZ} leads to
    \begin{multline}
        \begin{bmatrix}
            G_{\mathcal D}^\top \tilde Q G_{\mathcal D} & F_{\mathcal D}^\top & Y_{\set A}^\top\\
            F_{\mathcal D} & \mathbf 0 & \mathbf 0\\
            Y_{\set A} & \mathbf 0 & \mathbf 0
        \end{bmatrix}\begin{bmatrix}
            \xi^\star\\ \lambda^\star\\ \mu^\star_{\set A}
        \end{bmatrix} +\\ \bbm \hphantom{-}G_{\mathcal D}^\top \tilde H^\top\\-\theta_{2,\mathcal D}\\ \hphantom{-}\mathbf{0}_{n_{\set A}} \ebm x_0 + \bbm \hphantom{-}G_{\mathcal D}^\top \tilde Q c_{\mathcal D}\\ -\theta_{1,\mathcal D}\\ -\mathbf{1}_{n_{\set A}} \ebm = \mathbf{0}_{\bar D + \bar n_c + n_{\set A}}.
    \end{multline}
  
Solving for the primal solution $\xi^\star$ and the dual solutions $\lambda^\star$, $\mu_{\set A}^\star$ shows, by inspection, that they coincide with the form \eqref{eq:KKTsolution} if and only if $K_{(\set A)}$, $\kappa_{1,(\set A)}$, and $\kappa_{2,(\set A)}$ (cf.~\eqref{eq:notation_kkt_KA}--\eqref{eq:notation_kkt_k12}) are redefined to retain only their last $\bar n_c + n_{\set A}$ rows. Carrying the remaining steps in Prop.~\ref{prop:kkt_z} to arrive at \eqref{eq:affine_law}--\eqref{eq:cr} concludes the proof. \hfill\hfill\qed
\end{pf}
\vspace{-1em}
\subsection{Iterative updates}\label{subsec:accelerating}
\vspace{-1em}

In \eqref{eq:notation_kkt}, the matrices $Z_{(\mathbb A)}$, $T_{(\mathbb A)}^+$, and $K_{(\mathbb A)}^{-1}$ need not be recomputed from scratch when the active set $\mathbb A$ is updated. We next analyze the effect of an update $\bar{\set A} = \mathbb A \cup \{i\}$ on each of these matrices.

% Several improvements can be made in Algorithm~\ref{alg:computeCriticalRegion}. We need not compute the entire $Z_{(\set A)}$ matrix from scratch, mainly because Step~\ref{step:ZnullSpace} is costly for large matrices. Thus, for some $\bar{\set A} \supset \set A$, $Z_{(\bar{\set A})}$ can be obtained by updating a previously computed matrix $Z_{(\set A)}$; this result is captured in Lemma~\ref{lem:ZrecursiveUpdate}. Moreover, the null matrix has a particular structure that can be derived analytically, thus improving the computation time.

\begin{lem}\label{lem:null_update} Whenever $\bar{\set A} = \mathbb A \cup \{i\}$ holds, we have 
\begin{equation}\label{eq:Zupdate}
    Z_{(\bar{\set A})} = Z_{(\set A)} \Null\left(Y_i Z_{(\set A)}\right).
\end{equation}
\end{lem}
\vspace{-3em}
\begin{pf}
From $\bar{\set A} = \set A \cup \{i\}$, we have
\begin{equation*}
    Z_{(\bar{\set A})} 
    = \Null\!\left(\bbm F_{\mathcal D}^\top & Y_{\bar{\set A}}^\top \ebm^\top \right) 
    = \Null\!\left(\bbm F_{\mathcal D}^\top & Y_{\set A}^\top & Y_i^\top \ebm^\top \right).
\end{equation*}
Since $Z_{(\set A)} = \Null\bigl(\begin{bsmallmatrix} F_{\mathcal D}^\top & Y_{\set A}^\top \end{bsmallmatrix}^\top \bigr)$, any $z \in Z_{(\bar{\set A})}$ satisfies:  
i) $F_{\mathcal D} z = \mathbf{0}$,  
ii) $Y_{\set A} z = \mathbf{0}$,  
iii) $Y_i z = \mathbf{0}$.  
From (i) and (ii), it follows that $z \in Z_{(\set A)}$, so there exists $v$ with $z = Z_{(\set A)} v$. Substituting into (iii) yields the additional condition $Y_i Z_{(\set A)} v = 0$, i.e., $v \in \Null(Y_i Z_{(\set A)})$. Hence, $Z_{(\bar{\set A})} = Z_{(\set A)} \, \Null\!\left(Y_i Z_{(\set A)}\right)$, as in \eqref{eq:Zupdate}. 
 \hfill\hfill\qed
\end{pf}
\vspace{-1em}
While $\Null\!\left(Y_i Z_{(\set A)}\right)$ can be computed by standard methods (e.g., \texttt{QR} factorization), updating $K_{(\set A)}^{-1}$ is more efficient when a sparse representation is employed. The next result provides such a representation.

% \begin{lem}\label{lem:efficient_ker}
% Let $z\in \mathbb R^n$ be a row vector. Then, $\Null{\left(z\right)}$, has a basis $V\in \mathbb R^{n \times (n-1)}$ with entries
% \begin{equation*}
%     V_{ik} = \begin{cases}
%         1, & (i=k \land i<j) \lor (i=k-1 \land i>j),\\
%         -z_i/z_j, & i=j,\\
%         0, &\text{otherwise},
%     \end{cases}
% \end{equation*}
% with index $j$ chosen such that $z_j\neq 0$.
% \end{lem}
% \begin{pf}
% To show that the columns of $V$ form a basis for $\Null(z)$, it suffices to verify that each column is orthogonal to $z$. By construction, $z V_{:,k} = 0$ for all $k$, which completes the proof. \hfill\hfill \qed
% \end{pf}

\begin{lem}\label{lem:efficient_ker}
%Let $z\in\mathbb R^n$ be a nonzero row vector and choose an index $j$ with $z_j\neq 0$. Define the $n\times (n-1)$ matrix $V$ whose columns are
\hlight{R8-7}{Let $z\in\mathbb R^n$ be a nonzero vector and choose an index
$j\in\{1,\dots,n\}$, with $z_j\neq 0$. Define the $n\times (n-1)$ matrix
$V=(V_{ik})_{\substack{1\leq i\leq n\\1\leq k\leq n-1}}$ whose entries are
}
\begin{equation*} 
    V_{ik} = \begin{cases}
        1, & i=\sigma(k),\\
        -\frac{z_{\sigma(k)}}{z_j}, & i=j,\\
        0, &\text{otherwise},
    \end{cases}
\end{equation*}
where $\sigma(k)=k$, for $k<j$, and $\sigma(k)=k+1$, for $k\geq j$. Then, the columns of $V$ form a basis of $\Null\left(z^\top\right)$.
\end{lem}
\begin{pf}
Let $v^{(k)}=Ve_k$. By the definition of $V$, $v^{(k)}_{\sigma(k)}=1$, $v^{(k)}_j=-\frac{z_{\sigma(k)}}{z_j}$, and $v^{(k)}_i=0$, for $i\notin\{j,\sigma(k)\}$. Hence, $zv^{(k)}=z_{\sigma(k)}+z_j \left(-\frac{z_{\sigma(k)}}{z_j}\right)=0$, so every column $v^{(k)}\in\Null{\left(z^\top\right)}$. Thus, $\text{im}(V)\subseteq \Null{\left(z^\top\right)}$. Deleting the $j$-th row of $V$ results in a matrix whose columns are vectors of the canonical base of $\mathbb R^{n-1}$, which are linearly independent. %Therefore, any $v^{(k)}$ is linearly independent. 
Since $z\neq 0$, $\text{rank}\left(z\right)=1$, and, by rank-nullity, $\text{dim}\left( \Null{\left(z^\top\right)}\right)=n-1$. We have $n-1$ independent vectors in $\Null{\left(z^\top\right)}$, so they form a basis of $\Null{\left(z^\top\right)}$.
\hfill\hfill \qed
\end{pf}
\vspace{-1em}

We have now the prerequisite to compute the update of $K_{(\bar{\set A})}$ and of its inverse, $K_{(\bar{\set A})}^{-1}$.
\begin{prop}\label{prop:invKA}Whenever $\bar{\set A} = \mathbb A \cup \{i\}$ holds, we have 
    \begin{equation}
    \label{eq:ka_inverse_update}
        \mkern-12mu K_{(\bar{\set A})}^{-1} = \bigl[K_{({\set A})}^{-1}-K_{({\set A})}^{-1}U\mkern-4mu\left(I_2+WK_{({\set A})}^{-1}U\right)^{-1}\mkern-16mu WK_{({\set A})}^{-1}\bigr]P(j,\bar n_c + i)^\top,
    \end{equation}
where\vspace{-2em}
\begin{equation}
\label{eq:notation_uw}
    U = \left[\begin{array}{c:c} & \hphantom{-}v_1\\e_j & -1\\ &\hphantom{-}v_2\end{array}\right], \: W=\left[\begin{array}{c} Y_i\\\hdashline Z_{(\set A),j}^\top G_D^\top \tilde Q G_D\end{array}\right].
\end{equation} 
%and with $v=\bbm v_1^\top & v_2^\top\ebm^\top$, the $j$-th row from matrix $V$, chosen such that the null matrix update 
\hlight{R8-8}{Consider that $V$ is the matrix given by Lem.~\ref{lem:efficient_ker}. In particular,
$v=\begin{bsmallmatrix} v_1^\top & v_2^\top\end{bsmallmatrix}^\top$ denotes the $j$-th row of $V$, chosen so that the null-space update}
\vspace{-1em}
\begin{equation}
\label{eq:z_update}
    Z_{(\bar{\set A})} = Z_{(\set A)}V = Z_{(\set A)}\left[\begin{array}{c:c}I_{k_1}&\mathbf 0\\\hdashline v_1^\top& v_2^\top\\\hdashline \mathbf 0 & I_{k_2}\end{array}\right]
\end{equation}
holds; $P(j,\bar n_c + i)$ extracts the $j$-th row and inserts it at the $(\bar n_c + i)$-th position, shifting all subsequent rows, as needed.
\end{prop}
\vspace{-1em}
\begin{pf}
Computing $V$ as in Lem.~\ref{lem:efficient_ker} for $z = Y_i Z_{(\set A)}$ and applying Lem.~\ref{lem:null_update} yields the expression of $Z_{(\bar{\set A})}$ in \eqref{eq:z_update}. The update from $K_{({\set A})}$ to $K_{(\bar{\set A})}$ is twofold: (i) the $j$-th row is eliminated by a linear combination of the rows of $Z_{(\set A)}^\top$ with weightings given by $v$, and (ii) a new row $Y_i$ is inserted at the position $\bar n_c+i$. This can be modeled by first updating the $j$-th row and then moving it to the position $\bar n_c+i$. With the notation of \eqref{eq:notation_uw}, the update reads
\begin{equation}
    K_{(\bar{\set A})} = P(j,\bar n_c + i)\left(K_{({\set A})} + UW\right). 
\end{equation}
Then, noting that $P(j,\bar n_c + i)^{-1}= P(j,\bar n_c + i)^\top$ and applying the Woodbury matrix identity \hlight{R8-8}{gives \eqref{eq:ka_inverse_update}}, which concludes the proof. \hfill\hfill\qed
\end{pf}
\vspace{-2em}
For clarity, Fig.~\ref{fig:kkt_update} illustrates the updates $Z_{(\set A)} \rightarrow Z_{(\bar{\set A})}$ and $K_{(\set A)} \rightarrow K_{(\bar{\set A})}$. Patterned rectangles mark the rows modified during the update, while indices $1,2$ on matrices $Z_{(\set A)}^\top$, $Y_{\set A}$, and the vector $v$ denote the row blocks between which a single row is deleted or inserted.

% \begin{rem}
% By constructing the null space as described in Prop.~\ref{prop:efficient_ker}, we generate a sparse representation of $\Null(z)$, which is not guaranteed by using a classical factorization, such as the Singular Value Decomposition.
% \end{rem}
% \vspace{-3em}
\hlight{R9-2}[1cm]{\begin{rem}
\label{rem:low-rank-online}
Noteworthy,~\cite{valmorbida2023quadratic} also employs a low-rank update of the inverse KKT matrix, conceptually similar to~\eqref{eq:ka_inverse_update}, albeit in a polyhedral setting. However, the purpose of that update is to accelerate the online computation of the MPC control action, rather than to construct an explicit solution through critical-region enumeration.\eor    
\end{rem}}

\begin{rem} The invertibility of $K_{(\bar{\set A})}$ is determined by the factor $\bigl(I_2 + W K_{(\set A)}^{-1} U\bigr)$ in \eqref{eq:ka_inverse_update}. Since this matrix lies in $\mathbb R^{2\times 2}$, the singularity test is both efficient and numerically reliable. \eor
\end{rem}

% \clearpage
\begin{rem}
    Whenever $\bar{\set A} = \mathbb A \cup {i}$ holds, we have \cite{ben2003generalized}
    \begin{equation}
    \label{eq:ka_pinverse_update}
    T_{(\bar{\set A})}^{+} = S_k^\top\bbm T_{(\set A)}^+ -d b^\top\\ b^\top\ebm,
    \end{equation}
    where $d = T_{(\set A)}^+ Y_i^\top$, $c=\left(Y_i^\top - T_{(\set A)} d\right)$, and $b$ is a vector
    \begin{equation*}
    b = \begin{cases}
    \cfrac{c}{c^\top c}, & \sqrt{c^\top c}>\varepsilon,\\
    \cfrac{{T_{(\set A)}^{+\top}} d}{1+d^\top d}, &\text{otherwise}.
    \end{cases}
    \end{equation*}
    The term $S_k\in\mathbb R^{(n+1)\times (n+1)}$ is a column permutation that moves the last column to position $k$, i.e., $S_k = \left(\sum_{j=1}^{k-1}e_j e_j^\top + e_{n+1}e_{k}^\top + \sum_{j=k}^n e_j e_{j+1}^\top\right)$. We take $k=n_{F_{\mathcal D}}+\bigl|\{\ell\in\bar{\set A}:\ \ell\le i\}\bigr|$; lastly, $\varepsilon$ is a tolerance. \eor
    % maybe cite "https://bura.brunel.ac.uk/bitstream/2438/9127/2/Fulltext.pdf" or "https://arxiv.org/pdf/2005.07045"
\end{rem}

As shown in \cite{mihai2022explicit}, the active sets defining non-empty critical regions are embedded in the face lattice of the polyhedral feasible domain of the MPC problem. Here, owing to the simpler structure, we employ a lifted formulation in $\mathbb R^{\bar D}$ with variable $\xi$. According to \cite[Prop.~18]{jones2004equality}, there exists a surjective relation whereby each face of the feasible domain corresponds to a face in the lifted space (though not all faces of $\xi$ map to faces of $u$). We exploit this fact to propose a fast necessary check for the non-emptiness of a candidate set $\bar{\set A} = \set A \cup \{i\}$.
\begin{prop}\label{prop:quick_check}
Let $\set A$ be an active set such that $CR_{(\set A)}$ is non-empty. Define the corresponding active set of polyhedral constraints $\set A_{\mathrm{red}} \subset \{1,\ldots,q\}$ as
\begin{equation}
\label{eq:ared}
    \set A_{\mathrm{red}} = \{\, j : A_j \mathbf {u_{[0:N-1]}}_{(\set A)} = b_j + F_j x_0 \,\}.
\end{equation}
Then, for a candidate set $\bar{\set A} = \set A \cup \{i\}$, the following implication holds
\begin{equation}
\label{eq:quick_check}
    \Bigl(\not\exists j \text{ s.t. } \bar{\set A}_{\mathrm{red}} = \set A_{\mathrm{red}} \cup \{j\}\Bigr)
    \implies \bigl(CR_{(\bar{\set A})} = \emptyset \bigr).
\end{equation}
\end{prop}
\begin{pf}
For a candidate active set $\bar{\set A} = \set A \cup \{i\}$, we compute $\bar{\set A}_{\mathrm{red}}$ as in \eqref{eq:ared}. If there is no index $j$ such that $\bar{\set A}_{\mathrm{red}} = \set A_{\mathrm{red}} \cup \{j\}$, the candidate solution does not lie on a child face of the polyhedral representation. Since, under mild assumptions\footnote{We assume LICQ, which can be ensured via slight perturbations of the original constraints.}, the explicit solution is continuous and corresponds to a single index update in the active set, implication \eqref{eq:quick_check} follows, concluding the proof. \hfill\hfill\qed
\end{pf}
% \newpage
\section{Explicit MPC algorithm}\label{sec:algorithm}

We introduce Algorithm~\ref{alg:mpQP_CZ} to compute the solution of explicit MPC, adapting to the particular structure and properties of constrained zonotopes. The algorithm follows similar reasoning to that introduced in the previous work for the general polyhedral case (see \cite{mihai2022explicit}). 

\begin{algorithm}
\caption{Explicit MPC solution in the CZ case}\label{alg:mpQP_CZ}
\algrenewcommand\algorithmicindent{0.6em}
\begin{algorithmic}[1]
    \Require \hlight{R1-7}{The cost matrices $\tilde{Q}$ and $\tilde{H}$; the constrained zonotope matrices in \eqref{eq:notation_cz}}; the matrix $Y$ of the hypercube representation in \eqref{eq:mp-QP-CZ_c}.
    \Ensure The graph $\mathcal{L}'$ containing the solution
    \State Initialize $\mathcal{L}'$ with a node containing $\emptyset$, denoted as $n_\emptyset$
    \State Initialize a queue $\mathcal{Q}$ with $\emptyset$
    \While{$\mathcal{Q}$ is not empty}
        \State Extract active set $\set H$ from the queue $\mathcal{Q}$
        \label{step:get_parent}\State Compute $\set S_{(\set H)}$ as in \eqref{eq:indicesToKeep} \label{step:algoMS_step_toKeep}
        \label{step:sh}\State Initialize $\mathcal{G} \leftarrow \{\emptyset\}$
        \For{\textbf{each} $i$ in $\set S_{(\set H)}$}\label{step:startMinSets}
            \State Let $h' \leftarrow \set H$
            \State Make $h'(i) \leftarrow$ 0; $h'(i + D) \leftarrow$ 1\label{step:swap_1}
            \State $\mathcal G\leftarrow \{\mathcal G, h'\}$
            \State Make $h'(i) \leftarrow$ 1; $h'(i + D) \leftarrow$ 0\label{step:swap_2}
            \State $\mathcal G\leftarrow \{\mathcal G, h'\}$
        \EndFor\label{step:endMinSets}
        \State Let $n^{\set H}$ be a node for $\set H$ in $\mathcal L'$
        \For{$\textbf{each candidate set}\ \set{A} \in \mathcal{G}$}\label{step:forAinG}            \State\label{step:find_or_create} Find or create the node $n^{\set{A}}$ for $\set{A}$ in $\mathcal{L}'$
            \If{$n^{\set{A}}$ is newly created}
                \State Compute the critical region using Alg.~\ref{alg:computeCriticalRegion_recursive}
                \State $(L_{(\set A)}, \ell_{(\set A)}, \textit{flag}) \leftarrow \texttt{getCriticalRegion}(\set{A})$
                \If{\textit{flag} is \textbf{false}}\label{step:numerical_check}
                    \State \textbf{go to Step~\ref{step:forAinG}} [\textsc{numerical issues}]
                \EndIf
                \If{implication \eqref{eq:quick_check} does not hold}\label{step:quick_check}
                    \State \textbf{go to Step~\ref{step:forAinG}} [\textsc{quick check}]
                \EndIf
                \If{the polyhedral set $P(L_{(\set A)}, \ell_{(\set A)})$ is \texttt{empty}}\label{step:check_empty}
                    \State \textbf{go to Step~\ref{step:forAinG}} [\textsc{Chebyshev radius}]
                \EndIf
                \State Add the active set $\set{A}$ to the queue $\mathcal{Q}$
                \State Add the edge $(n^{\set H}, n^{\set{A}})$ to the graph $\mathcal{L}'$
            \EndIf
            \State Add the edge $(n^{\set H}, n^{\set{A}})$ to $\mathcal{L}'$ if not already added
        \EndFor
    \EndWhile
    \State \Return the solution graph $\mathcal L'$
\end{algorithmic}
\end{algorithm}

Alg.~\ref{alg:mpQP_CZ} enumerates nodes from the face lattice of the feasible domain \eqref{eq:domain}, each corresponding to a non-empty critical region. The resulting structure is a tree in which every node is defined by its active set, together with the associated primal/dual solution and critical region (plus auxiliary data required for internal updates, e.g., $Z_{(\set A)}$, $K_{(\set A)}^{-1}$, and $T^+_{(\set A)}$). The edges are indexed by the constraints which become active in the transition from parent to child. 

Although the procedure is conceptually simple, its efficiency depends critically on three aspects: i) generation of candidate active sets for a given parent node; ii) reuse of previously computed information in determining the critical region; and iii)  a hierarchy of emptiness checks and stopping conditions to avoid unnecessary computations. These aspects are discussed in detail next.

First, since active sets are in bijection with the faces of the feasible domain, each potential child of a parent node ($\set H$ from step~\ref{step:get_parent}) is either a neighboring face of a higher order or the empty set. In the polyhedral case, this enumeration requires facet–vertex operations \cite{mihai2022explicit}, which are computationally burdensome. In contrast, the constrained zonotope description of \eqref{eq:domain} enables explicit enumeration in steps~\ref{step:startMinSets}--\ref{step:endMinSets}.  \hlight{R8-3}{Noteworthy, step~\ref{step:find_or_create} exploits the tree structure of $\mathcal L'$ to determine whether the current node is already contained within.}

Assume that the inequalities $|\xi|\leq \mathbf 1_{\bar D}$ in \eqref{eq:domain} are ordered such that the $i$-th inequality corresponds to $\xi_i \leq 1$ and the $(i+D)$-th to $-\xi_i \leq 1$. We then extract the indices for which neither\footnote{Both constraints cannot be simultaneously active, since this would require $\xi_i = 1$ and $\xi_i = -1$, respectively.} constraint in the pair $(i,i+D)$ is active
\begin{equation}\label{eq:indicesToKeep}
    \set S_{(\set A)} = \left\{\, i \in \{1,\dots,D\} : a_i \veebar a_{i+D}=0,\ i \in \set A \,\right\}.
\end{equation}
Here, the active set $\set A = \{a_1,a_2,\dots,a_{2D}\}$ is encoded with logical values, as $a_i \in \{0,1\}$, where $a_i=1$ if the $i$-th constraint is active. The exclusive OR is defined as
\begin{equation}\label{eq:xor}
    x \veebar y = \begin{cases}
        1, & x \neq y,\\
        0, & x = y.
    \end{cases}
\end{equation}
Enumerating over each $i \in \set S_{(\set A)}$ and activating either the $i$-th or $(i+D)$-th constraint, as done in steps~\ref{step:startMinSets}--\ref{step:endMinSets}, generate all child facets of the face described by $\set A$.

Secondly, Alg.~\ref{alg:computeCriticalRegion_recursive} details the computation of the critical region in the constrained-zonotope case. For brevity, we only highlight the improvements over \cite{stoican2024computing}, namely the iterative construction of the matrices $Z_{(\bar{\set A})}$, $K_{(\bar{\set A})}$, and $T^+_{(\bar{\set A})}$ (Steps~\ref{step:update_ZA},~\ref{step:update_KA} and~\ref{step:update_TA}).
\begin{algorithm}
\caption{(\texttt{getCriticalRegion}) Compute the critical region with CZ using iterative updates}\label{alg:computeCriticalRegion_recursive}
\begin{algorithmic}[1]
    \Require The current active set $\bar{\set{A}}$ of the form $\bar{\set{A}} = \set A \cup \{i\}$, a tolerance $\varepsilon$
    \Ensure Critical region $(L_{(\bar{\set A})}, \ell_{(\bar{\set A})})$ and a flag \textit{success}
    \State Let $D \leftarrow$ number of columns in $G_D$
    \State Initialize $L_{(\bar{\set A})} \leftarrow \emptyset$, $\ell_{(\bar{\set A})} \leftarrow \emptyset$, \textit{success} $\leftarrow$ \textbf{true}
    \State Retrieve $Z_{(\set A)}$, $K_{(\set A)}^{-1}$, and $T^+_{(\set A)}$ from memory\label{step:memory}
    \State Compute $Z_{(\bar{\set A})} = Z_{(\set A)} \Null\left(Y_i Z_{(\set A)}\right)$ as in Lemma~\ref{lem:null_update}\label{step:update_ZA}
    \If{$Z_{(\bar{\set A})}^\top G_D^\top Q G_D Z_{(\bar{\set A})}$ is not positive definite}
        \State \textit{success} $\leftarrow$ \textbf{false}  [\textsc{numerical infeasibility}]\label{step:num_infeas}
        \State \Return $(L_{(\bar{\set A})}, \ell_{(\bar{\set A})}, \textit{success})$
    \EndIf
    \State Let $k\leftarrow\set A\veebar \bar{\set A}$ and compute $T^+_{(\bar{\set A})}$ as in \eqref{eq:ka_pinverse_update}\label{step:update_TA}
    \State Determine matrices $U$, $V$ as in \eqref{eq:notation_uw}
    \If{$\min_{\lambda \in \Lambda\!\left(I_2 + V K^{-1}_{({\set A})} U\right)} |\lambda| \leq \varepsilon$}
        \State \textit{success} $\leftarrow$ \textbf{false} [\textsc{singular matrix detected}]\label{step:singular_mx}
        \State \Return $(L_{(\bar{\set A})}, \ell_{(\bar{\set A})}, \textit{success})$
    \EndIf
    \State Compute $K^{-1}_{(\bar{\set A})}$ as in Prop.~\ref{prop:invKA}\label{step:update_KA}
    \State Store  $Z_{(\bar{\set A})}$, $K_{(\bar{\set A})}^{-1}$, and $T^+_{(\bar{\set A})}$  in memory
    \State Compute $S_{(\bar{\set A})}$ and $s_{(\bar{\set A})}$ with \eqref{eq:notation_kkt_SA}, \eqref{eq:notation_kkt_sA}
    \State Extract $L_{(\bar{\set A})}$ and $\ell_{(\bar{\set A})}$ using \eqref{eq:cr}
    \State \Return $(L_{(\bar{\set A})}, \ell_{(\bar{\set A})}, \textit{success})$
\end{algorithmic}
\end{algorithm}
\vspace{-1em}

\hlight{R8-10}{In Step~\ref{step:memory}, the term ``memory'' denotes the per-node cache associated with the solution tree. More precisely, for each accepted active set $\set A$, the matrices $Z_{(\set A)}$, $K^{-1}_{(\set A)}$, and $T^+_{(\set A)}$ are stored and reused when constructing child nodes. This cache is used only during the offline construction of the explicit solution. Its dominant storage cost is $O(n_{\mathrm{nodes}} \bar D^2)$, mainly due to storing the matrices $K^{-1}_{(\set A)}$ for all accepted nodes.}

Thirdly, we clarify the stopping conditions in Alg.~\ref{alg:mpQP_CZ}. Since the full emptiness test is computationally expensive, we adopt a hierarchical approach. If Alg.~\ref{alg:computeCriticalRegion_recursive} returns \textbf{false} (e.g., due to numerical issues or lack of an inverse), we stop early at step~\ref{step:numerical_check}. If the necessary condition in Prop.~\ref{prop:quick_check} fails at step~\ref{step:quick_check}, we also terminate. Only at step~\ref{step:check_empty} we do perform the full emptiness test. Among the available strategies, our practical implementation computes the Chebyshev radius \cite{boyd2004convex} of the candidate region and applies a threshold-based rule to discard regions with small radius.

\begin{figure*}[!ht]
    \includegraphics[width=.9\textwidth]{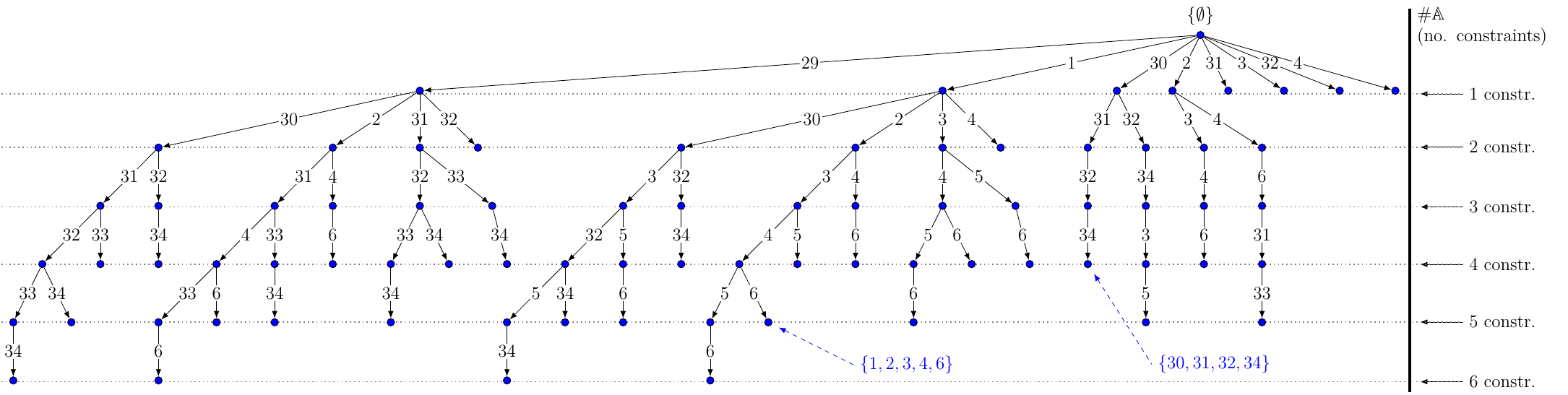}
    \caption{Tree structure obtained from Alg.~\ref{alg:mpQP_CZ}, applied to dynamics \eqref{eq:example_dynamics}, for prediction horizon $N=4$.}
    \label{fig:tree}
\end{figure*}
\section{Numerical simulation}\label{sec:simulation}
This section presents a comparison of the algorithms introduced previously. For clarity, we label the results as follows: Alg.~\ref{alg:mpQP_CZ} is labeled \texttt{CZ}, and the improvements described in Section~\ref{subsec:accelerating} are labeled \texttt{CZ (Iterative)} and \texttt{CZ (Iterative | Quick)}. \hlight{R8-11}[1cm]{}We test the methods on an example with the state vector in $\mathbb{R}^4$ given as
\hlight{}{
\begin{align}
\label{eq:example_dynamics}
    A_d = \begin{bmatrix}
    \hphantom{-}0.84 & \hphantom{-}0.02 & \hphantom{-}0.00 & -0.02\\
    \hphantom{-}0.01 & \hphantom{-}0.88 & -0.03            & -0.02\\
    \hphantom{-}0.04 & -0.08            & \hphantom{-}0.86 & -0.02\\
   -0.09            & -0.02            & -0.02            & \hphantom{-}0.88
\end{bmatrix}
,
    B_d = \begin{bmatrix}
   -0.10            & \hphantom{-}0.03\\
   -0.07            & -0.07\\
   -0.27            & \hphantom{-}0.13\\
    \hphantom{-}0.01 & -0.00
\end{bmatrix}.
\end{align}
}
The associated cost matrices are $S=Q=I_{4}$ and $R=0.1\cdot I_{2}$, with state and input constraints $|x_k|\leq 10\cdot \mathbf{1}_n$ and $|u_k|\leq \mathbf{1}_m$. 
% All algorithms are run in \texttt{MATLAB R2024b} on a laptop with \texttt{Intel 9} and 32 GB of RAM, on \texttt{Windows 11}.
\rev{All algorithms are run in MATLAB on a dedicated Ubuntu workstation with an Intel hybrid processor, with eight performance cores (16 logical CPUs). All methods have the same MATLAB and BLAS/LAPACK configuration, and our OpenMP kernels use $T\in\{1,8\}$ threads. In the MPT3 and YALMIP case, any internal parallelism is managed by their own back-ends, not by the CZ thread setting.}
\hlight{\parbox{1cm}{R1-1 R3-1 R8-13 R8-15 R9-4}}[-.5cm]{The critical pieces of code (tree search/update and critical-region construction) have been implemented as batch \texttt{mex} calls (C++ code that accepts MATLAB data structures and which, after compilation, are called transparently from MATLAB).}

To illustrate the solution, we apply Alg.~\ref{alg:mpQP_CZ} to the dynamics \eqref{eq:example_dynamics} with prediction horizon $N=4$, yielding the tree structure shown in Fig.~\ref{fig:tree}. Each edge is labeled with the index of the constraint that becomes active, so the active set of any node can be obtained by tracing a path from the root (empty set) to that node (two examples are highlighted). Although multiple paths may exist depending on the order in which indices are appended, the result is a tree, since Alg.~\ref{alg:mpQP_CZ} terminates the search once a link is established.

\begin{figure}[!ht]
\centering
\precompiledplot{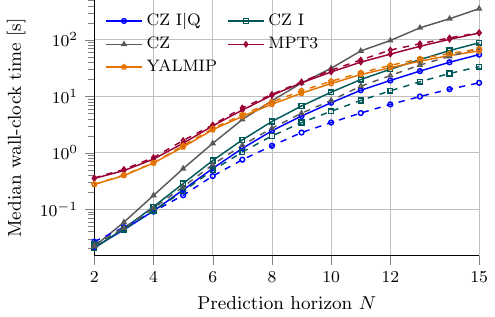}{15625636}{10001075}{0}{0}{0}
\caption{\rev{Median computation time over ten repetitions vs. prediction horizon $N$: solid lines, $T=1$; dashed lines, $T=8$.}}
\label{fig:computation_time_vs_N}
\end{figure}

% Next, Fig.~\ref{fig:computation_time_vs_N} shows the computation time of each algorithm variant as a function of the prediction horizon. Each data point is averaged over 10 runs, with the corresponding values reported in Table~\ref{tab:computation_times_paper}. \hlight{R9-3}{We compare our algorithm with the state-of-the-art solvers, namely YALMIP,~\cite{lofberg2004yalmip}, and MPT3,~\cite{herceg2013multi}}. As \hlight{}{a first conclusion, both iterative variants of our algorithm are consistently faster than YALMIP and MPT3\footnote{\label{foot:yalmip_vs_mpt3}\textcolor{black}{Note that YALMIP exploits the constrained zonotopic representation \eqref{eq:mp-QP-CZ} of the multi-parametric problem and that MPT3 solves it in the standard polyhedral form \eqref{eq:mp-QP}.}}. Noteworthy, at $N=12$ we observe a $32\%$ reduction.}\hlight{\parbox{1cm}{R1-1 R3-1 R8-13 R8-15 R9-4}}[-.5cm]{Second, in all cases, the iterative quick variant is the best performer and the baseline CZ (non-iterative) degrades the fastest. This is inline with expectations and confirms the advantages provided by, first using iterative constructions and, second, pre-processing candidate critical regions.} 
\rev{Next, Fig.~\ref{fig:computation_time_vs_N} compares each algorithm variant with YALMIP,~\cite{lofberg2004yalmip}, and MPT3,~\cite{herceg2013multi}, for $N=2,\ldots,15$}
\footnote{\label{foot:yalmip_vs_mpt3}\textcolor{black}{Note that YALMIP exploits the constrained zonotopic representation \eqref{eq:mp-QP-CZ} of the multi-parametric problem and that MPT3 solves it in the standard polyhedral form \eqref{eq:mp-QP}.}}
\rev{. Each point represents the median wall-clock time over ten repetitions, and Table~\ref{tab:computation_times_paper} provides the $T=8$ values. All iterative variants perform better than the references at $T=8$. At $T=1$, the iterative quick variant remains faster, but the iterative variant is slower than YALMIP for $N\geq13$.}

% \begin{figure}[!ht]
%     \centering
%     \resizebox{\columnwidth}{!}{%
%         \input{./images/comp_time_vs_N_updated.tikz}%
%     }
%     \caption{Computation time vs. prediction horizon $N$.}
%     \label{fig:computation_time_vs_N}
% \end{figure}
% \vspace{-1em}
% {\small\begin{table}[!ht]
% \centering%\color{blue}
% \caption{Average computation times for the considered methods.}
% \label{tab:computation_times_paper}
% \begin{tabular}{cccccc}
% \toprule
% $N$ &
% CZ I|Q &
% CZ I&
% CZ &
% MPT3 &
% YALMIP \\
% \midrule
% 2  & 0.03   & 0.02   & 0.06   & 0.59    & 0.51   \\
% 3  & 0.07   & 0.09   & 0.37   & 1.12    & 1.13   \\
% 4  & 0.23   & 0.31   & 1.54   & 2.45    & 2.56   \\
% 5  & 0.70   & 1.00   & 5.65   & 5.53    & 5.82   \\
% 6  & 1.91   & 2.83   & 17.18  & 12.22   & 12.54  \\
% 7  & 4.86   & 7.53   & 45.46  & 23.45   & 23.64  \\
% 8  & 10.50  & 16.60  & 93.54  & 46.96   & 45.70  \\
% 9  & 22.88  & 34.99  & 199.04 & 107.83  & 95.21  \\
% 10 & 76.37  & 98.16  & 620.76 & 250.71  & 207.86 \\
% 11 & 275.04 & 316.22 & -    & 540.10  & 463.33 \\
% 12 & 586.24 & 659.39 & -    & 1237.56 & 859.55 \\
% \bottomrule
% \end{tabular}
% \end{table}
% }

\begin{table}[!ht]
\centering
\caption{\rev{Median computation times [s] over ten repetitions, $T=8$.}}
\label{tab:computation_times_paper}
\begingroup\revcolor\small
\setlength{\tabcolsep}{3.6pt}
\renewcommand{\arraystretch}{0.94}
\pgfplotstabletypeset[
  col sep=comma, columns={N,O5,O6,O7,O9,O10}, column type=r,
  fixed, precision=3, fixed zerofill,
  columns/N/.style={column name=$N$,precision=0,fixed zerofill=false},
  columns/O5/.style={column name={CZ I$|$Q}},
  columns/O6/.style={column name={CZ I}},
  columns/O7/.style={column name={CZ}},
  columns/O9/.style={column name={MPT3}},
  columns/O10/.style={column name={YALMIP}},
  every head row/.style={before row=\toprule,after row=\midrule},
  every last row/.style={after row=\bottomrule},
]{\benchmarkresults/paper1/p1_time_T8.csv}
\endgroup
\end{table}

\begin{table*}[!t]
\centering
\caption{\rev{Computation times [s] for ten random systems (one run per scenario). CZ I$|$Q: constrained-zonotope method with iterative updates and quick tests; EO: enumeration only; CC: completeness check and graph completion.\protect\footnotemark\ $n,m,N$: state dimension, input dimension and prediction horizon; $T$: CZ OpenMP thread count. ``regions'': partition size after completion; ``missed'': percentage added. MPT3 and YALMIP use the $T=1$ configuration (similar performance to $T=8$ in their case).}}
\label{tab:scalability}
\begingroup\revcolor\small
\renewcommand{\arraystretch}{0.96}
\begin{tabular*}{\textwidth}{@{\extracolsep{\fill}}rrrrrrrrrrrr@{}}
\toprule
\multicolumn{4}{c}{System} & \multicolumn{2}{c}{Partition}
 & \multicolumn{2}{c}{CZ I$|$Q, $T=1$}
 & \multicolumn{2}{c}{CZ I$|$Q, $T=8$}
 & \multicolumn{2}{c}{Reference solvers, $T=1$}\\
\cmidrule(lr){1-4}\cmidrule(lr){5-6}\cmidrule(lr){7-8}\cmidrule(lr){9-10}\cmidrule(lr){11-12}
\# & $n$ & $m$ & $N$ & regions & \shortstack{missed\\{[\%]}}
 & EO [s] & EO+CC [s]& EO [s]& EO+CC [s]& MPT3 [s]& YALMIP [s]\\
\midrule
1 & 4 & 1 & 15 & 5\,019 & 33.0 & 15.14 & 58.71 & 6.38 & 49.95 & 260.7 & 141.6 \\
2 & 4 & 2 & 7 & 20\,564 & 45.8 & 21.62 & 161.5 & 12.93 & 152.8 & 1\,359.9 & 1\,328.7 \\
3 & 4 & 2 & 15 & 16\,101 & \(<\)0.1 & 131.5 & 173.3 & 41.21 & 83.00 & 257.9 & 137.6 \\
4 & 4 & 2 & 15 & 18\,016 & 13.2 & 172.4 & 291.0 & 55.02 & 173.6 & 480.6 & 334.3 \\
5 & 4 & 2 & 15 & 51\,577 & 0.4 & 533.7 & 726.4 & 176.9 & 369.6 & 1\,234.2 & 861.1 \\
6 & 4 & 2 & 11 & 57\,982 & 6.4 & 267.7 & 513.2 & 110.2 & 355.7 & 1\,183.6 & 949.4 \\
7 & 5 & 2 & 4 & 19\,761 & 1.0 & 23.59 & 66.70 & 16.58 & 59.70 & 371.5 & 1\,573.5 \\
8 & 5 & 2 & 6 & 30\,455 & 14.6 & 43.79 & 204.1 & 25.37 & 185.7 & 3\,137.5 & 2\,888.6 \\
9 & 5 & 2 & 14 & 33\,285 & 5.6 & 378.1 & 652.1 & 105.6 & 379.6 & 1\,198.3 & 632.8 \\
10 & 5 & 2 & 10 & 39\,816 & 1.4 & 189.3 & 403.1 & 67.90 & 281.7 & 1\,328.6 & 723.8 \\
\bottomrule
\end{tabular*}
\endgroup
\end{table*}

% \hlight{\parbox{1cm}{R1-1 R3-1 R3-8 R9-5}}{Scalability remains a challenging issue, as it does for all active-set enumeration strategies. To the best of our knowledge, no analytical bound exists that relates the problem dimensions ($n$) and prediction horizon length ($N$) to the resulting number of critical regions. Furthermore, the latter depends not only on these quantities, but also on the cost matrices and on the shape and magnitude of the constraints. While not comprehensive, we provide in Table~\ref{tab:scalability} some representative examples. We selected random systems with states and inputs in $n\in \{2, \ldots, 6\}$, $m\in \{1,2\}$ for prediction horizons from $N\in \{2, \ldots, 12\}$ and to which we applied our own \texttt{CZ (Iterative)} and the MPT3 and YALMIP variants. Further details (tuning matrices, constraints, dynamics) are found in the \texttt{/inputs} folder of our public repository.}

\rev{For $N=15$ and $T=8$, we observe that CZ (Iterative | Quick) completes in $17.3$\,s, whereas MPT3 and YALMIP require $134.2$\,s and $69.1$\,s, respectively. This yields speedups of approximately $7.7$ over MPT3 and $4.0$ over YALMIP, relative to their respective $T=8$ runtimes.}

\rev{Scalability remains a challenging issue, because critical region counts depend on the dimensions, horizon, cost matrices and constraints. Table~\ref{tab:scalability} shows ten demanding random system cases using \texttt{CZ (Iterative | Quick)} at $T=\{1,8\}$ and the reference solvers (similar performance at either $T=1$ or $T=8$).}

\begin{figure}[!ht]
\centering
\precompiledplot{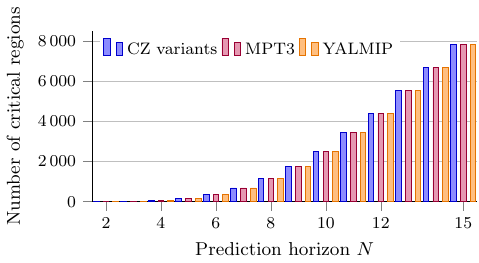}{15124361}{8295085}{0}{0}{0}
\caption{\rev{Number of critical regions vs. prediction horizon $N$, $T=8$. The three CZ variants have identical counts.}}
\label{fig:nregions_vs_N}
\end{figure}

\rev{Fig.~\ref{fig:nregions_vs_N} shows the total number of regions at $T=\{1,8\}$. The CZ variants agree throughout the tested prediction horizons. At $N=15$, our variants and YALMIP identify $7819$ regions, while MPT3 identifies $7814$. We observe that the discrepancies relative to CZ remain below $0.07\%$, which likely results from differences in the numerical tolerances used in the emptiness tests.}

\begin{figure}[!ht]
\centering
\precompiledplot{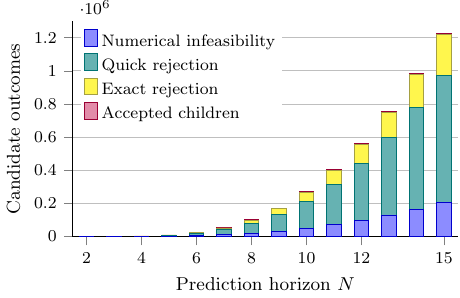}{14514505}{9385215}{0}{0}{0}
\caption{\rev{Candidate outcomes for the iterative quick variant through $N=15$. Counts are identical at all tested thread levels.}}
\label{fig:exit_conditions}
\end{figure}

\rev{Fig.~\ref{fig:exit_conditions} shows the outcomes for the iterative quick variant. Notably, at $N=15$, the quick test rejects $\approx75\%$ of the $1\,024\,824$ candidates passing the numerical checks. The numerical rejections are related to infeasibility or a singular matrix (see Steps~\ref{step:num_infeas} and~\ref{step:singular_mx} of Alg.~\ref{alg:computeCriticalRegion_recursive}).}

\begingroup
\interfootnotelinepenalty=10000
\footnotetext{\rev{Graph completion tests points near critical region facets and adds a missing neighboring region when it is discovered. The completeness check and graph completion are run serially with $T=8$, and their execution time is added to the enumeration time for both CZ thread settings.}}
\endgroup

\section{Conclusions}
\label{sec:con}
\vspace{-1em}
This paper presented a reinterpretation of explicit MPC in which the feasible domain is represented as a constrained zonotope. The multi-parametric problem was formulated in the lifted generator space and solved using second-order optimality conditions. Computational efficiency was enhanced through low-rank matrix updates and an analytic enumeration of candidate active sets naturally yielded the explicit solution in tree form. 
% \hlight{\parbox{1cm}{R1-1 R3-1 R3-8 R9-5}}{Numerical results demonstrated improved computational performance over conventional polyhedral formulations for the benchmark example considered and exhibited encouraging scalability trends across a collection of randomly generated dynamical systems.}
\rev{We observed improved computational performance for the example when using the medians of ten repetitions with the number of threads $T=1$ and $T=8$, evaluated over prediction horizons $N=2,\ldots,15$.}

The code used to generate the results is available at \url{www.gitlab.com/msstefan/constrained-zonotope-empc/-/tree/paper-1-iterative-updates}.

\onecolumn
\appendix

\section{Benchmark analysis}

This appendix reports an extended benchmark; the results of Section~\ref{sec:simulation} are taken from it.
% ==========================================================================
\subsection{Benchmark platform and control of multithreading}\label{sec:platform}

All benchmarks run on a dedicated Ubuntu workstation with an Intel hybrid processor, 8 performance (P) cores with two hardware threads each and 4 efficiency (E) cores (20 hardware threads), and 32\,GB of RAM. We made every source of parallelism explicit:
\begin{itemize}[leftmargin=1.5em]
\item the MATLAB process is pinned to the 16 logical CPUs of the 8 P-cores
  (\texttt{taskset}); the E-cores are never used;
\item MATLAB's own computational threads and the BLAS/LAPACK library are
  limited to one thread\\ (\texttt{maxNumCompThreads = 1}, MKL with one
  thread; at \(T=1\) MATLAB is started with \texttt{-singleCompThread});
\item our parallel kernels use OpenMP with exactly \(T\) threads, set per
  call and in the environment (\texttt{OMP\_NUM\_THREADS},
  \texttt{OMP\_THREAD\_LIMIT}, \texttt{OMP\_DYNAMIC=false}); idle OpenMP
  threads sleep (\texttt{OMP\_WAIT\_POLICY=passive});
\item the campaign is run for \(T\in\{1,2,4,8,16\}\); \(T=1\) is a
  strictly serial run of every method. With \(T\le 8\) every thread can have
  its own P-core; \(T=16\)
  necessarily places two threads on each P-core (simultaneous
  multithreading).
\end{itemize}
All methods are run for \(\{N=2,\dots,15\}\), ten repetitions per configuration, in the same MATLAB session layout. Every relevant variable of the CZ enumeration (no. of candidates, rejections of each kind, number of tests, regions) is identical for all thread levels, so the parallel execution does
not change what is computed. 

\textbf{Completeness of the partition.} Whether an enumeration finds \emph{all} critical regions is a complex question. Our method is a combinatorial, top-down exploration of the active sets in which a child differs from its parent by a single index; the geometric methods, such as the one in MPT3, step across the facets of the regions found so far. In both cases degeneracy can break the link between
neighboring regions, either primal (more
constraints active than can be linearly independent) and dual
(active constraints with zero multipliers) degeneracy may make a region reachable only through a change of more than one index or break the facet-to-facet property. No enumeration
scheme is immune to this in general.

So as not to modify the enumeration of Alg.~\ref{alg:mpQP_CZ}, we append an \emph{a posteriori} check and geometric completion step: every facet of every critical region is flipped, i.e.\ a point just beyond the facet is tested; it must either lie outside the feasible parameter set or inside a neighboring region of the partition. A point that satisfies
neither reveals a gap; the optimal active set at that point is then computed
and the missing region is added, and the check is repeated until no gap
remains. For the example of Section~\ref{sec:simulation} and all horizons \(N=2,\dots,15\),
for all three CZ methods, no gap was found and no
region had to be added (Table~\ref{tab:coverage}). The check is a verification of the result, not part of the algorithm: its time is \emph{not} included in any of the reported totals and is listed separately in Table~\ref{tab:coverage}. MPT3 and YALMIP were not passed through the
check; they return the same number of regions up to a few units (at \(N=15\): \(7\,811\)--\(7\,817\) for MPT3 for MPT3 and \(7\,818\)--\(7\,819\) for YALMIP,
against \(7\,819\)).

\begin{table}[!ht]
\centering\small
\IfFileExists{\pdir/p1_coverage.csv}{%
\pgfplotstabletypeset[
  columns={N,regions,facets,gapsFound,regionsAdded,checkSeconds},
  columns/regions/.style={column name={\shortstack{regions\\checked}}},
  columns/facets/.style={column name={\shortstack{facets\\flipped}}},
  columns/gapsFound/.style={column name={\shortstack{gaps\\found}}},
  columns/regionsAdded/.style={column name={\shortstack{regions\\added}}},
  columns/checkSeconds/.style={column name={\shortstack{check\\time [s]}},precision=2,zerofill},
  % columns/checkPct_O5/.style={column name={\shortstack{vs CZ\\(It.\(\mid\)Q) [\%]}}},
  % columns/checkPct_O6/.style={column name={\shortstack{vs CZ\\(It.) [\%]}}},
  % columns/checkPct_O7/.style={column name={\shortstack{vs CZ\\{[\%]}}}},
  skip rows between index={0}{4},
]{\pdir/p1_coverage.csv}}{\fbox{missing p1\_coverage.csv}}
\caption{A posteriori completeness check of the CZ partitions (\(N\ge 6\)).}
\label{tab:coverage}
\end{table}

\begin{figure}[!ht]
\centering
\precompiledplot{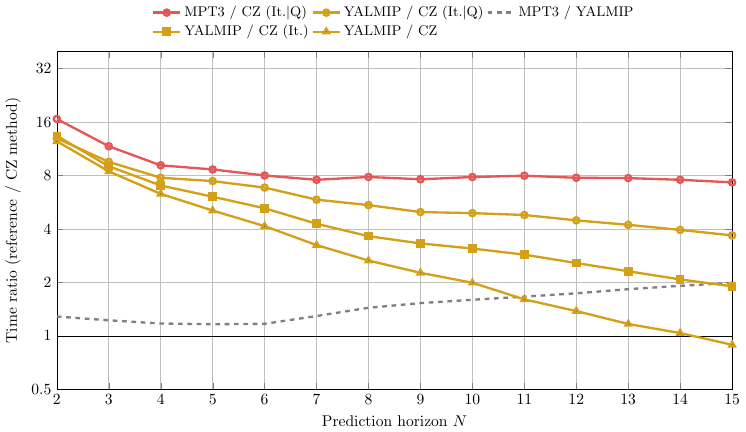}{23526090}{13627134}{0}{0}{0}
\caption{Ratios of the best times (each method at its own best thread
level). Above one, the CZ method is faster.}
\label{fig:ratios}
\end{figure}

\textbf{MPT3 and YALMIP and the thread limit.} Our thread setting controls our own OpenMP kernels. MPT3 and YALMIP do not use them; whatever parallelism they have comes from their own back-ends and is not governed by our setting in the same way. This is confirmed by the fact that their wall-clock time hardly changes with \(T\): at
\(N=15\), MPT3 needs \(132.1\), \(126.9\), \(130.1\), \(134.2\) and
\(187.7\)\,s at \(T=1,2,4,8,16\), YALMIP \(63.8\), \(68.1\), \(68.5\),
\(69.1\) and \(73.1\)\,s. Thus, it is fair to run them at every thread level and to
compare against their \emph{best} time, which is what
Fig.~\ref{fig:ratios} does.

% A remark on CPU time: the ratio CPU-time/wall-time grows almost linearly with
% \(T\) for all methods, including MPT3 (\(1.0\), \(1.8\), \(3.2\), \(6.3\), \(10.3\) at
% \(T=1,2,4,8,16\), \(N=15\)), although their wall time does not decrease. This ratio counts
% every thread that holds a core, also while it waits for work, and is not a
% measure of useful parallel work; we no longer use it (it was shown in our
% previous reply) and compare wall-clock times only.

% ==========================================================================
\subsection{Implementation details}\label{sec:implementation}

The implementation follows Alg.~\ref{alg:mpQP_CZ} exactly: same candidate generation, same low-rank (iterative) updates, same quick and exact emptiness tests, same
order. The details below do not change the computed partitions.

\textbf{Compiled kernels (MEX files).} The construction of a critical region (CR) from its parent by the low-rank update, the quick test and the exact emptiness
test (an LP solved with DAQP, compiled into the same library) are C++/C MEX functions.

\textbf{OpenMP.} All children of an expanded node are independent of each
other: each one is obtained from the same parent data by its own low-rank
update, and is tested independently. Now, one MEX call handles the whole batch of children of a node, and its loop is distributed over \(T\) OpenMP threads. The exact tests of a batch are run the same way, with one DAQP workspace per thread. The advantages of OpenMP over MATLAB-level parallelism (\texttt{parfor}, process pools) are significant and visible in the subsequent results. 

\textbf{The quick test inside the region construction.} The quick test of Section~\ref{subsec:accelerating} needs only the affine parametrization of the candidate (the
quantities \(L\) and \(l\) of the solution) and the parent data. These are the first quantities produced by the low-rank update. Thus, the test is executed inside the region construction, right after they are available, and a rejected child is abandoned there: the update of the
pseudo-inverse, the dual (multiplier) rows and the construction of the half-space representation of its region are never computed. Algorithmically, the result is identical to implementing the emptiness test separately.

\textbf{Deferring every expensive step until it is needed.} The same
principle is applied along the whole pipeline of a candidate, ordered by
increasing cost:
\begin{center}
\precompiledplot{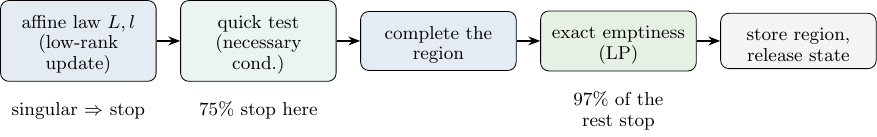}{27675004}{4310772}{0}{-0.075}{0}
\end{center}
Only the candidates that survive a stage pay for the next one. After a node
has been expanded, the data needed only to generate its children (inverse
and pseudo-inverse factors, screening data) are released, which keeps the
memory footprint proportional to the stored solution.

% ==========================================================================
\subsection{Enumeration details}\label{sec:enumeration}

\begin{figure}[!ht]
\centering
\precompiledplot{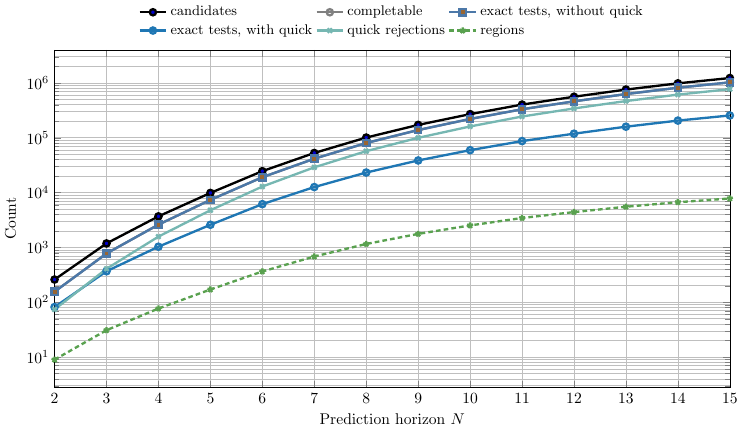}{23454410}{13577961}{0}{0}{0}
\caption{Growth of the enumeration quantities with the horizon. All curves
grow at a similar rate; the quick test lowers the number of exact tests by a
factor that rises from \(3.1\) at \(N=6\) to \(4.0\) at \(N=15\).}
\label{fig:funnel}
\end{figure}

The total time is the result of several mechanisms, each with its own
growth with \(N\). Table~\ref{tab:funnel} and Fig.~\ref{fig:funnel} illustrate this behavior, for:

\begin{itemize}[leftmargin=1.5em]
\item \emph{Candidate regions.} Each parent node generates one candidate per added admissible index: \(1.23\) million candidates at \(N=15\), for \(7\,819\) regions. Only \(0.64\%\) of the candidates become regions; the cost of an enumeration is dominated by the candidates that are rejected.
\item \emph{Numerical infeasibility} (\(16.6\%\) of the candidates at
  \(N=15\)). The low-rank update detects a singular KKT system immediately;
  these candidates cost a fraction of a region construction.
\item \emph{Quick rejections} (\texttt{CZ (Iterative | Quick)} only). \(75.1\%\) of
  the remaining candidates are rejected by the quick test, inside the
  construction, before their region is completed.
\item \emph{Exact tests.} With the quick test \(255\,064\) LPs are solved at
  \(N=15\) instead of \(1\,024\,824\) (\(4\times\) fewer); \(97\%\) of them
  still find an empty region.
\end{itemize}

% --- funnel table --------------------------------------------------------
\begin{table}[!ht]
\centering\small
\IfFileExists{\pdir/p1_funnel.csv}{%
\pgfplotstabletypeset[
  columns={N,candidates,numInfeasible,O5_quickRejected,O5_exactTests,O6_exactTests,regions,quickRejectPct,yieldPct},
  /pgf/number format/fixed, /pgf/number format/precision=0,
  columns/N/.style={column name=\(N\)},
  columns/candidates/.style={column name={candidates}},
  columns/numInfeasible/.style={column name={\shortstack{numerically\\infeasible}}},
  columns/O5_quickRejected/.style={column name={\shortstack{quick\\rejected\(^{a}\)}}},
  columns/O5_exactTests/.style={column name={\shortstack{exact tests\\with quick\(^{a}\)}}},
  columns/O6_exactTests/.style={column name={\shortstack{exact tests\\without quick\(^{b}\)}}},
  columns/regions/.style={column name={regions}},
  columns/quickRejectPct/.style={column name={\shortstack{quick\\rej.\ [\%]}},fixed,precision=1,zerofill},
  columns/yieldPct/.style={column name={\shortstack{yield\\{[\%]}}},fixed,precision=2,zerofill},
  skip rows between index={0}{4},
]{\pdir/p1_funnel.csv}}{\fbox{missing p1\_funnel.csv}}
\caption{What happens to the candidates (\(N\ge 6\)). \(^{a}\) CZ
(Iterative \(\mid\) Quick); \(^{b}\) CZ (Iterative) and CZ. ``Quick rej.''
is the share of the completable candidates (candidates minus numerically
infeasible) rejected by the quick test; ``yield'' is regions/candidates.}
\label{tab:funnel}
\end{table}

\FloatBarrier
% ==========================================================================
\subsection{Time consumption}\label{sec:time}

\begin{figure}[!ht]
\centering
\precompiledplot{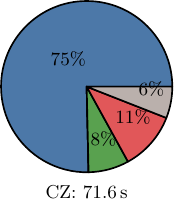}{5459984}{6473261}{0}{-0.1}{0}\hfill
\precompiledplot{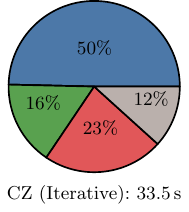}{5948328}{6660041}{0}{0}{0}\hfill
\precompiledplot{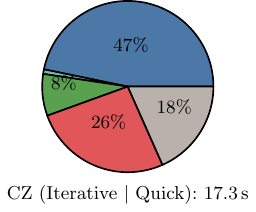}{8070042}{6660041}{0}{0}{0}\\[0.5em]
\precompiledplot{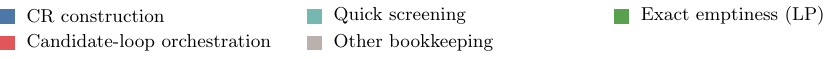}{26253634}{1833015}{0}{-0.075}{0}
\caption{Wall-time decomposition at \(N=15\), \(T=8\) (total time under each
pie). Each step of the method removes work from the dominant category.}
\label{fig:pies}
\end{figure}

We partition the measured wall-clock time of each CZ method into disjoint categories: CR construction (the batched low-rank updates; with \(T>1\) it
includes the quick test executed inside it), the quick test, the exact emptiness tests, the candidate-loop orchestration in
MATLAB, and the remaining bookkeeping (child generation, queue, tree insertion and lookup, residual). Figure~\ref{fig:pies} shows the three methods at \(N=15\), \(T=8\).

\begin{figure}[!ht]
\centering
\precompiledplot{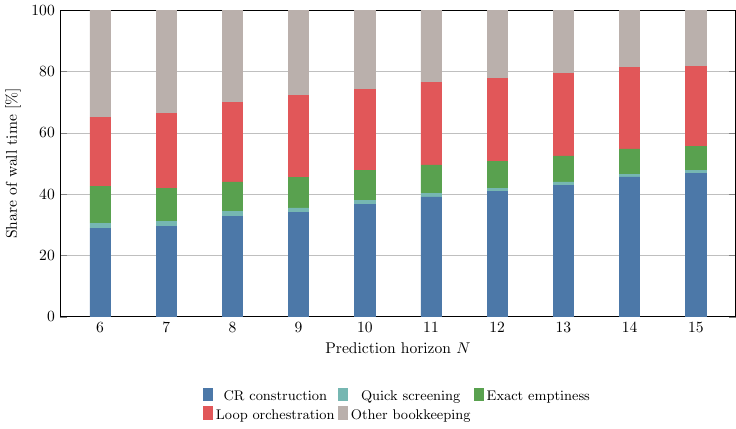}{23235585}{13523931}{0}{0}{0}
\caption{Wall-time decomposition of CZ (Iterative \(\mid\) Quick) at
\(T=\Tmain\). The two red and grey blocks are MATLAB-level orchestration and
bookkeeping.}
\label{fig:bars}
\end{figure}

Reading the pies from left to right shows the contribution of each element of the method:
\begin{itemize}[leftmargin=1.5em]
\item \texttt{CZ}: the region construction from scratch dominates
  (\(54.0\)\,s, \(75\%\)).
\item \texttt{CZ (Iterative)}: the low-rank updates make the construction
  \(3.2\times\) cheaper (\(16.7\)\,s); the exact tests (\(5.3\)\,s) and the
  orchestration become relevant.
\item \texttt{CZ (Iterative | Quick)}: the construction halves again
  (\(8.1\)\,s), because \(75\%\) of the children are abandoned before their
  region is completed, and the exact tests drop to \(1.4\)\,s.
\end{itemize}
MATLAB ancillary operations are significant. The candidate loop and the bookkeeping take about \(7.7\)\,s, i.e.\ \(44\%\) of the \(17.3\)\,s (Fig.~\ref{fig:bars}): the loop passes through each of the \(1.23\) million
candidates and spends about \(4\,\mu\)s per candidate on creating and copying small structures, growing cell arrays,
indexing, plus the insertion into the search tree. 

\FloatBarrier
% ==========================================================================
\subsection{Effect of the number of threads}\label{sec:threads}

\begin{figure}[!ht]
\centering
\precompiledplot{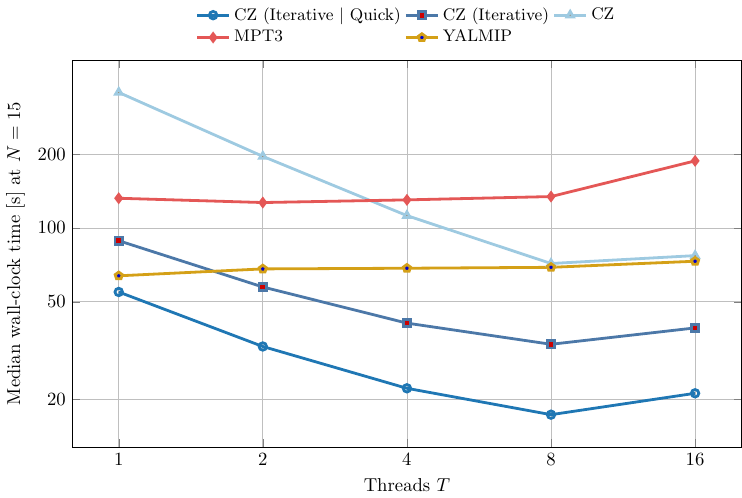}{23423705}{15710481}{0}{0}{0}
\caption{Wall-clock time versus number of threads at \(N=\Nthr\). Only the CZ
methods use the OpenMP kernels; MPT3 and YALMIP are shown at every level for
reference.}
\label{fig:threads}
\end{figure}

Table~\ref{tab:threads} and Fig.~\ref{fig:threads} give the times at
\(N=\Nthr\) for every thread level; Table~\ref{tab:components} splits the CZ times into
the parallel kernels and the rest.

\begin{itemize}[leftmargin=1.5em]
\item The parallel part scales well: from \(T=1\) to \(T=8\) the CR construction
  of \texttt{CZ} goes from \(334.2\) to \(54.0\)\,s (\(6.2\times\) on 8
  threads, a parallel efficiency of about \(77\%\)). For
  \texttt{CZ (Iterative | Quick)} the kernels as a whole (CR construction,
  quick and exact tests) go from \(45.9\) to \(9.6\)\,s (\(4.8\times\)); their
  tasks are smaller, so the fixed cost of each parallel batch weighs more.
  (At \(T=1\) the quick test executed inside the construction is reported
  under ``tests'', for \(T>1\) under ``CR'' in Table~\ref{tab:components}.)
\item The pure MATLAB part does not scale: the orchestration and bookkeeping take
  \(8.9\)\,s at \(T=1\) and \(7.6\)--\(7.7\)\,s from \(T=2\) to \(T=8\) for
  \texttt{CZ (Iterative | Quick)}, \(12.0\) and \(11.7\)--\(12.2\)\,s for
  \texttt{CZ}. From \(T=1\) to \(T=8\) the
  overall gain is \(3.2\times\) for \texttt{CZ (Iterative | Quick)} and
  \(2.6\times\) for \texttt{CZ (Iterative)}, whose MATLAB-level share grows
  from \(16\%\) and \(13\%\) of the serial time to \(44\%\) and \(34\%\) at
  \(T=8\), and \(5.0\times\) for \texttt{CZ}, whose time is dominated by the
  parallel construction (MATLAB share \(3\%\) serially).
\item \(T=16\) is slower than \(T=8\) for every method: the two threads of a
  P-core share its execution units and caches, so the parallel part gains
  nothing, while the serial part slows down (the orchestration of
  \texttt{CZ (Iterative | Quick)} grows from \(7.7\) to \(10.4\)\,s).
\item \texttt{CZ (Iterative | Quick)} is faster than both references at every
  thread level and horizon, the serial run included. \texttt{CZ (Iterative)}
  is faster than both from \(T=2\) on; serially it remains faster than MPT3
  but is slower than YALMIP from \(N=13\) (\(88.7\) against \(63.8\)\,s at
  \(N=15\)). The basic \texttt{CZ} method is slower than YALMIP from \(N=8\)
  and than MPT3 from \(N=10\) serially, from \(N=10\) and \(N=13\) with
  \(T=2\), and for \(T\ge 4\) only at the largest horizons and only against YALMIP.
\end{itemize}

\begin{table}[!ht]
\centering\small
\IfFileExists{\pdir/p1_threads_N\Nthr.csv}{%
\pgfplotstabletypeset[
  columns={T,O5,O6,O7,O9,O10,O5_speedup,O6_speedup,O7_speedup},
  columns/T/.style={column name=\(T\)},
  columns/O5/.style={column name={\shortstack{CZ\\(It.\(\mid\)Q)}},fixed,precision=2,zerofill},
  columns/O6/.style={column name={\shortstack{CZ\\(It.)}},fixed,precision=2,zerofill},
  columns/O7/.style={column name={CZ},fixed,precision=2,zerofill},
  columns/O9/.style={column name={MPT3},fixed,precision=2,zerofill},
  columns/O10/.style={column name={YALMIP},fixed,precision=2,zerofill},
  columns/O5_speedup/.style={column name={\shortstack{speed-up\\CZ (It.\(\mid\)Q)}},fixed,precision=2,zerofill},
  columns/O6_speedup/.style={column name={\shortstack{speed-up\\CZ (It.)}},fixed,precision=2,zerofill},
  columns/O7_speedup/.style={column name={\shortstack{speed-up\\CZ}},fixed,precision=2,zerofill},
]{\pdir/p1_threads_N\Nthr.csv}}{\fbox{missing p1\_threads\_N\Nthr.csv}}
\caption{Median wall-clock time [s] at \(N=\Nthr\) per thread level, and
speed-up with respect to the serial run (\(T=1\)).}
\label{tab:threads}
\end{table}

\begin{table}[!ht]
\centering\small
\IfFileExists{\pdir/p1_components_N\Nthr.csv}{%
\pgfplotstabletypeset[
  columns={T,O5_total,O5_cr,O5_tests,O5_glue,O5_glue_pct,O7_total,O7_cr,O7_tests,O7_glue,O7_glue_pct},
  columns/T/.style={column name=\(T\)},
  columns/O5_total/.style={column name={total},fixed,precision=1,zerofill},
  columns/O5_cr/.style={column name={CR},fixed,precision=1,zerofill},
  columns/O5_tests/.style={column name={tests},fixed,precision=1,zerofill},
  columns/O5_glue/.style={column name={MATLAB},fixed,precision=1,zerofill},
  columns/O5_glue_pct/.style={column name={[\%]},fixed,precision=0,column type/.add={}{|}},
  columns/O7_total/.style={column name={total},fixed,precision=1,zerofill},
  columns/O7_cr/.style={column name={CR},fixed,precision=1,zerofill},
  columns/O7_tests/.style={column name={tests},fixed,precision=1,zerofill},
  columns/O7_glue/.style={column name={MATLAB},fixed,precision=1,zerofill},
  columns/O7_glue_pct/.style={column name={[\%]},fixed,precision=0},
  every head row/.style={
    before row={\toprule & \multicolumn{5}{c|}{CZ (Iterative \(\mid\) Quick)} & \multicolumn{5}{c}{CZ}\\},
    after row=\midrule},
]{\pdir/p1_components_N\Nthr.csv}}{\fbox{missing p1\_components\_N\Nthr.csv}}
\caption{Parallel kernels versus MATLAB-level work at \(N=\Nthr\) [s].
``Tests'' = quick + exact emptiness tests; ``MATLAB'' = candidate-loop
orchestration and bookkeeping (everything outside the kernels).}
\label{tab:components}
\end{table}

\FloatBarrier
% ==========================================================================
\subsection{Comparison with MPT3 and YALMIP}\label{sec:comparison}

\begin{figure}[!ht]
\centering
\precompiledplot{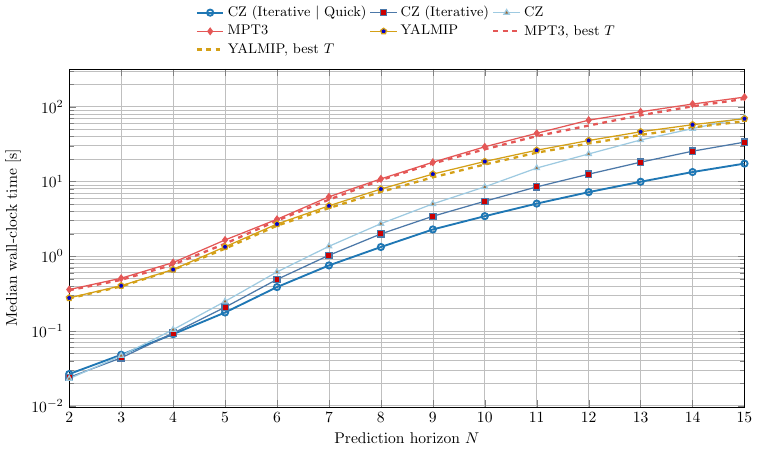}{23917260}{14180056}{0}{0}{0}
\caption{Median wall-clock time at \(T=\Tmain\) (solid). Dashed: MPT3 and
YALMIP at their best thread level for each \(N\). The vertical axis is
logarithmic.}
\label{fig:times}
\end{figure}

\begin{figure}[!ht]
\centering
\precompiledplot{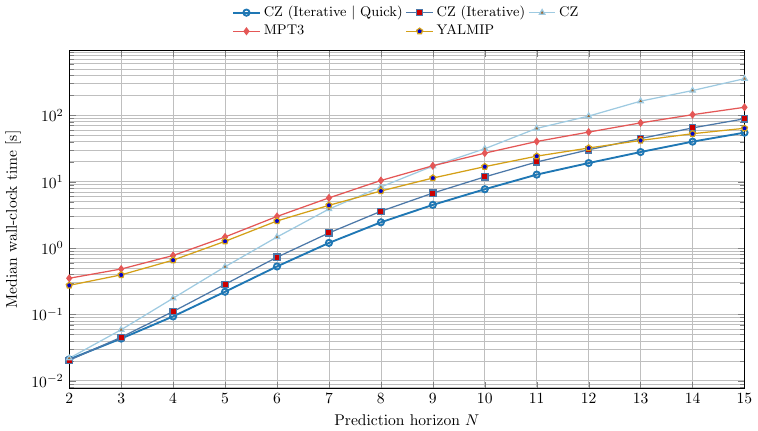}{23917260}{13597692}{0}{0}{0}
\caption{Median wall-clock time with one thread
for every method (\(T=1\)).}
\label{fig:times1}
\end{figure}

The two reference methods are both needed since:
\begin{itemize}[leftmargin=1.5em]
\item \emph{MPT3} works on the classical polyhedral formulation. The ratio
  MPT3/\texttt{CZ (Iterative | Quick)} stays between \(7.3\) and \(8.0\) for
  \(N=8,\dots,15\) (best against best): at this problem size the gain is stable. For the $T=1$ comparison it is \(2.4\times\) at \(N=15\).
\item \emph{YALMIP} uses the constrained-zonotope formulation \eqref{eq:mp-QP-CZ}, and is itself faster than MPT3, increasingly so with \(N\)
  (\(1.4\times\) at \(N=8\), \(2.0\times\) at \(N=15\), dashed curve). This independently confirms that the formulation pays off. Still, our enumeration adds a further factor on top: \(3.7\times\) for \texttt{CZ (Iterative |
  Quick)} and \(1.9\times\) for \texttt{CZ (Iterative)} at \(N=15\), best  against best. In the serial ($T=1$) comparison the margin of
  \texttt{CZ (Iterative | Quick)} over YALMIP is \(1.2\times\) at \(N=15\).
\end{itemize}

% --------------------------------------------------------------------------
% Effort per critical region (computed from p1_time_T<T>.csv and
% p1_funnel.csv; each method divided by its own number of regions)
% --------------------------------------------------------------------------
\newcommand{\percrcol}[3]{% #1 table, #2 optimizer index, #3 new column: ms per region
  \pgfplotstablecreatecol[create col/expr={1000*\thisrow{O#2}/\thisrow{O#2_regions}}]{#3}{#1}}
\newcommand{\percrcurve}[4]{% #1 = T, #2 = optimizer index, #3 = style, #4 = legend ({} = none)
  \IfFileExists{\pdir/p1_time_T#1.csv}{%
    \addplot+[#3,line width=1.5pt] table[x=N,col sep=comma,
      y expr={1000*\thisrow{O#2}/\thisrow{O#2_regions}}]{\pdir/p1_time_T#1.csv};
    \if\relax\detokenize{#4}\relax\else\addlegendentry{#4}\fi}{}}

\begin{figure}[!ht]
\centering
\precompiledplot{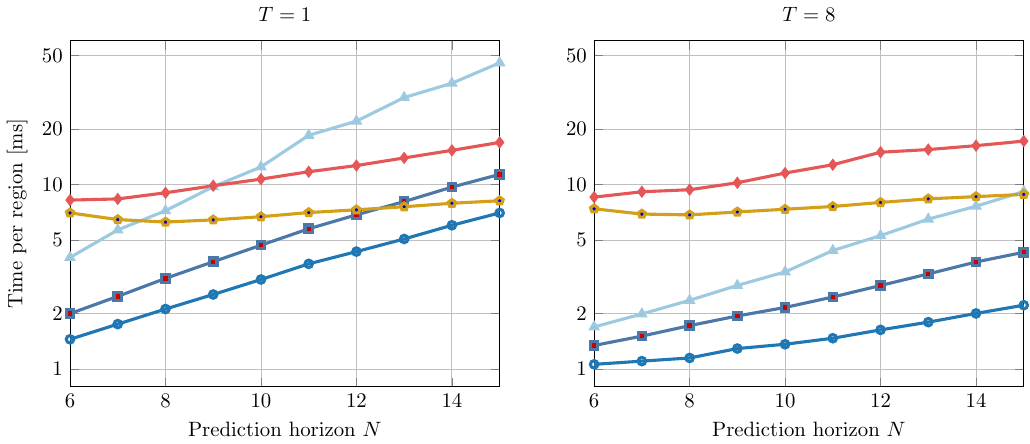}{32443520}{13991926}{0}{0}{0}\\[0.3em]
\precompiledplot{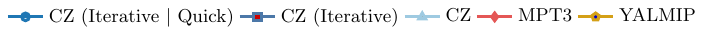}{22299202}{1064956}{0}{0}{0}
\caption{Median wall-clock time per critical region, serial run (left) and
\(T=\Tmain\) (right). The vertical axis is logarithmic.}
\label{fig:percr}
\end{figure}

\textbf{Effort per critical region.} The five methods return partitions of
the same size up to a few regions, so the time divided by the number of
regions of each method compares the cost of one region and shows how this
cost grows with the problem (Table~\ref{tab:percr}, Fig.~\ref{fig:percr}).
\begin{itemize}[leftmargin=1.5em]
\item With 8 threads, \texttt{CZ (Iterative | Quick)} needs \(2.2\)\,ms per
  region at \(N=15\), against \(17.2\)\,ms for MPT3 and \(8.8\)\,ms for
  YALMIP; serially, \(7.0\)\,ms against \(16.9\) and \(8.2\)\,ms.
\item The cost per region of the CZ methods grows with \(N\)
  (\texttt{CZ (Iterative | Quick)}: from \(1.1\) to \(2.2\)\,ms between
  \(N=6\) and \(N=15\) with 8 threads, from \(1.5\) to \(7.0\)\,ms serially).
  The reason is visible in Table~\ref{tab:percrwork}: the number of
  candidates per region grows linearly with \(N\), from \(67\) to \(157\),
  and each candidate is a larger problem. The quick test keeps the number of
  exact tests per region at \(17\)--\(33\), against \(52\)--\(131\) without
  it.
\item The cost per region of MPT3 doubles over the same range (\(8.5\) to
  \(17.2\)\,ms with 8 threads), that of YALMIP is almost constant
  (\(6.8\)--\(8.8\)\,ms). In the serial run the margin of
  \texttt{CZ (Iterative | Quick)} over YALMIP therefore narrows with \(N\),
  in agreement with Fig.~\ref{fig:times1}.
\end{itemize}

\begin{table}[!ht]
\centering\small\setlength{\tabcolsep}{4pt}
\IfFileExists{\pdir/p1_time_T\Tmain.csv}{\IfFileExists{\pdir/p1_time_T1.csv}{%
\pgfplotstableread[col sep=comma]{\pdir/p1_time_T\Tmain.csv}\pcrMain
\pgfplotstableread[col sep=comma]{\pdir/p1_time_T1.csv}\pcrOne
\percrcol\pcrMain{5}{m5}\percrcol\pcrMain{6}{m6}\percrcol\pcrMain{7}{m7}%
\percrcol\pcrMain{9}{m9}\percrcol\pcrMain{10}{m10}%
\percrcol\pcrOne{5}{s5}\percrcol\pcrOne{6}{s6}\percrcol\pcrOne{7}{s7}%
\percrcol\pcrOne{9}{s9}\percrcol\pcrOne{10}{s10}%
\pgfplotstablecreatecol[copy column from table={\pcrOne}{s5}]{s5}\pcrMain
\pgfplotstablecreatecol[copy column from table={\pcrOne}{s6}]{s6}\pcrMain
\pgfplotstablecreatecol[copy column from table={\pcrOne}{s7}]{s7}\pcrMain
\pgfplotstablecreatecol[copy column from table={\pcrOne}{s9}]{s9}\pcrMain
\pgfplotstablecreatecol[copy column from table={\pcrOne}{s10}]{s10}\pcrMain
\pgfplotstabletypeset[
  columns={N,O5_regions,m5,m6,m7,m9,m10,s5,s6,s7,s9,s10},
  /pgf/number format/fixed, /pgf/number format/precision=2,
  /pgf/number format/zerofill,
  columns/N/.style={column name=\(N\),precision=0,zerofill=false},
  columns/O5_regions/.style={column name={regions},precision=0,zerofill=false,
    column type/.add={}{|}},
  columns/m5/.style={column name={\shortstack{CZ\\(It.\(\mid\)Q)}}},
  columns/m6/.style={column name={\shortstack{CZ\\(It.)}}},
  columns/m7/.style={column name={CZ}},
  columns/m9/.style={column name={MPT3}},
  columns/m10/.style={column name={YALMIP},column type/.add={}{|}},
  columns/s5/.style={column name={\shortstack{CZ\\(It.\(\mid\)Q)}}},
  columns/s6/.style={column name={\shortstack{CZ\\(It.)}}},
  columns/s7/.style={column name={CZ}},
  columns/s9/.style={column name={MPT3}},
  columns/s10/.style={column name={YALMIP}},
  every head row/.style={
    before row={\toprule & & \multicolumn{5}{c|}{\(T=\Tmain\)} & \multicolumn{5}{c}{\(T=1\)}\\},
    after row=\midrule},
  skip rows between index={0}{4},
]\pcrMain}{\fbox{missing p1\_time\_T1.csv}}}{\fbox{missing p1\_time\_T\Tmain.csv}}
\caption{Median wall-clock time per critical region [ms] (\(N\ge 6\)), each
method divided by its own number of regions. }
\label{tab:percr}
\end{table}

\begin{table}[!ht]
\centering\small
\IfFileExists{\pdir/p1_funnel.csv}{%
\pgfplotstableread[col sep=comma]{\pdir/p1_funnel.csv}\pcrWork
\pgfplotstablecreatecol[create col/expr={\thisrow{candidates}/\thisrow{regions}}]{candPer}\pcrWork
\pgfplotstablecreatecol[create col/expr={\thisrow{O5_quickRejected}/\thisrow{regions}}]{quickPer}\pcrWork
\pgfplotstablecreatecol[create col/expr={\thisrow{O5_exactTests}/\thisrow{regions}}]{exactQPer}\pcrWork
\pgfplotstablecreatecol[create col/expr={\thisrow{O6_exactTests}/\thisrow{regions}}]{exactPer}\pcrWork
\pgfplotstabletypeset[
  columns={N,regions,candPer,quickPer,exactQPer,exactPer},
  /pgf/number format/fixed, /pgf/number format/precision=1,
  /pgf/number format/zerofill,
  columns/N/.style={column name=\(N\),precision=0,zerofill=false},
  columns/regions/.style={column name={regions},precision=0,zerofill=false},
  columns/candPer/.style={column name={\shortstack{candidates\\per region}}},
  columns/quickPer/.style={column name={\shortstack{quick rejections\\per region\(^{a}\)}}},
  columns/exactQPer/.style={column name={\shortstack{exact tests\\per region\(^{a}\)}}},
  columns/exactPer/.style={column name={\shortstack{exact tests\\per region\(^{b}\)}}},
  skip rows between index={0}{4},
]\pcrWork}{\fbox{missing p1\_funnel.csv}}
\caption{Work per critical region (\(N\ge 6\)); algorithmic counts,
identical for all thread levels. \(^{a}\) CZ (Iterative \(\mid\) Quick);
\(^{b}\) CZ (Iterative) and CZ.}
\label{tab:percrwork}
\end{table}

% ==========================================================================
% Section on the randomly generated systems (campaign lti25).
% Insert before "7. Outlook" and renumber the Outlook as Section 8.
% ==========================================================================
\FloatBarrier
\section{Randomly generated systems}\label{sec:random}

Table~\ref{tab:scalability} gives representative computation times for random systems. Here we extend that comparison to \(25\) random systems and add the completeness check. Unlike the example of Section~\ref{sec:simulation}, these systems do have partitions in which the enumeration misses regions, so this section also reports how many regions the check had to add and what it costs.

\textbf{Methodology.}
\begin{itemize}[leftmargin=1.5em]
\item The systems come from MATLAB's \texttt{rss} (random stable
  continuous-time model with \(n\) states and \(m\) inputs), discretized with \(T_s=0.1\)\,s, with the weights \(Q=I_n\), \(R=0.1\,I_m\) and the constraints \(|x_k|\le 10\cdot\mathbf 1_n\), \(|u_k|\le\mathbf 1_m\) of the example. Twenty systems were generated for each class \((n,m)\in\{(2,1),(3,1),(4,1),(4,2),(5,2)\}\) (file \texttt{input/lti\_systems.mat} of the repository); the first five of each class are used (indices \(1\)--\(5\), \(21\)--\(25\), \(41\)--\(45\),
  \(61\)--\(65\), \(81\)--\(85\); the index is given in
  Table~\ref{tab:lti}).
\item \texttt{CZ (Iterative | Quick)} is run with \(T=1\) and \(T=8\)
  threads, MPT3 and YALMIP with \(T=1\) (their times do not depend on \(T\),
  Section~\ref{sec:platform}). Horizons \(N=2,\dots,15\), one run per scenario.
\item Once a method needs more than \(1\,000\)\,s for a system at some \(N\), the larger horizons of that system are not run for that method.
\item The completeness check and completion of Section~\ref{sec:platform} (facet flipping;
  each gap point gives the optimal active set and its region is added; at
  most \(10\) rounds of check and repair) is run once per system and horizon. It is single threaded code (MATLAB with
  single-threaded compiled routines: \texttt{cdd}, DAQP, \texttt{linprog}),
  so its time is added to the enumeration time at both thread levels. 
\end{itemize}
We report for the CZ method the time of the enumeration only (EO, as in the rest of this appendix and in Section~\ref{sec:simulation}) and the time of the enumeration plus the
completeness check and completion (EO+CC). EO+CC returns a complete partition
(small variations in total CR numbers are marked with \(\dagger\) in Table~\ref{tab:lti}) and is the
fair counterpart of MPT3 and YALMIP; EO is the time of Alg.~\ref{alg:mpQP_CZ}. Table~\ref{tab:lti} gives representative cases (extending Table~\ref{tab:scalability}), Table~\ref{tab:ltiagg} the statistics over all
scenarios and Table~\ref{tab:lticov} the completeness results.

\begin{table}[!ht]
\centering\footnotesize\setlength{\tabcolsep}{3.2pt}
\begin{tabular}{rrrr|rr|rr|rr|rr}
\toprule
 & & & & & & \multicolumn{2}{c|}{CZ (It.\(\mid\)Q), \(T=1\)}
 & \multicolumn{2}{c|}{CZ (It.\(\mid\)Q), \(T=8\)} & MPT3 & YALMIP\\
\# & \(n\) & \(m\) & \(N\) & regions & \shortstack{missed\\{[\%]}}
 & EO [s] & EO+CC [s] & EO [s] & EO+CC [s] & [s] & [s]\\
\midrule
4 & 2 & 1 & 15 & 95 & 0.0 & 0.29 & 0.48 & 0.24 & 0.43 & 1.09 & 0.77 \\
\midrule
22 & 3 & 1 & 15 & 1\,023 & 41.6 & 1.79 & 6.34 & 0.93 & 5.48 & 13.69 & 7.55 \\
23 & 3 & 1 & 15 & 409 & 2.0 & 1.11 & 2.28 & 0.63 & 1.80 & 6.54 & \(\star\) \\
\midrule
41 & 4 & 1 & 15 & 2\,103 & 2.3 & 9.24 & 14.47 & 3.56 & 8.79 & 33.25 & 16.26 \\
43 & 4 & 1 & 15 & 5\,019 & 33.0 & 15.14 & 58.71\(^{\dagger}\) & 6.38 & 49.95\(^{\dagger}\) & 260.7 & 141.6 \\
\midrule
61 & 4 & 2 & 7 & 20\,564 & 45.8 & 21.62 & 161.5\(^{\dagger}\) & 12.93 & 152.8\(^{\dagger}\) & 1\,359.9 & 1\,328.7 \\
61 & 4 & 2 & 11 & 75\,527\(^{b}\) & \(\star\) & 383.6 & \(\star\) & 160.3\(^{a}\) & \(\star\) & -- & -- \\
62 & 4 & 2 & 15 & 16\,101 & \(<\)0.1 & 131.5 & 173.3 & 41.21 & 83.00 & 257.9 & 137.6 \\
63 & 4 & 2 & 15 & 18\,016 & 13.2 & 172.4 & 291.0\(^{\dagger}\) & 55.02 & 173.6\(^{\dagger}\) & 480.6 & 334.3 \\
64 & 4 & 2 & 15 & 51\,577 & 0.4 & 533.7 & 726.4 & 176.9 & 369.6 & 1\,234.2 & 861.1 \\
65 & 4 & 2 & 11 & 57\,982 & 6.4 & 267.7 & 513.2\(^{\dagger}\) & 110.2 & 355.7\(^{\dagger}\) & 1\,183.6 & 949.4 \\
65 & 4 & 2 & 14 & 115\,449\(^{b}\) & \(\star\) & 970.4 & \(\star\) & 338.1\(^{a}\) & \(\star\) & -- & -- \\
\midrule
81 & 5 & 2 & 4 & 19\,761 & 1.0 & 23.59 & 66.70\(^{\dagger}\) & 16.58 & 59.70\(^{\dagger}\) & 371.5 & 1\,573.5 \\
81 & 5 & 2 & 5 & 59\,272 & 1.3 & 99.05 & 295.3\(^{\dagger}\) & 62.32 & 258.5\(^{\dagger}\) & \(\star\) & -- \\
81 & 5 & 2 & 7 & -- & -- & \(\star\) & -- & \(\star\) & -- & -- & -- \\
82 & 5 & 2 & 6 & 30\,455 & 14.6 & 43.79 & 204.1\(^{\dagger}\) & 25.37 & 185.7\(^{\dagger}\) & 3\,137.5 & 2\,888.6 \\
83 & 5 & 2 & 7 & 40\,888 & 7.4 & 99.43 & 316.4\(^{\dagger}\) & 49.85 & 266.8\(^{\dagger}\) & 5\,999.0 & -- \\
84 & 5 & 2 & 14 & 33\,285 & 5.6 & 378.1 & 652.1\(^{\dagger}\) & 105.6 & 379.6\(^{\dagger}\) & 1\,198.3 & 632.8 \\
85 & 5 & 2 & 10 & 39\,816 & 1.4 & 189.3 & 403.1\(^{\dagger}\) & 67.90 & 281.7\(^{\dagger}\) & 1\,328.6 & 723.8 \\
85 & 5 & 2 & 13 & 103\,641\(^{b}\) & \(\star\) & 970.6 & \(\star\) & 278.9\(^{a}\) & \(\star\) & -- & -- \\
\bottomrule
\end{tabular}
\caption{Representative random systems.
\#: index of the system in \texttt{input/lti\_systems.mat}. Regions: size of
the partition after the completion. Missed: share of these regions that the
enumeration did not find (added by the completion). EO: enumeration only;
EO+CC: enumeration plus completeness check and completion (the check time is
the same for both thread levels). MPT3 and YALMIP: \(T=1\).
\(\star\): failed (MPT3 \#81, \(N=5\): stopped after \(2\)\,h; YALMIP
\#23: error at every horizon; CZ \#81, \(N=7\): numerical failure of the
exact emptiness test, at both thread levels; CZ with \(T=8\), \#61
\(N=11\), \#65 \(N=14\), \#85 \(N=13\): process killed for lack of memory
during the check). --: not run, because the same method exceeded the
\(1\,000\)\,s budget, or failed, at a smaller \(N\) for this system.
\(^{a}\)~Enumeration time read from the run log (the enumeration had
finished when the check ran out of memory). \(^{b}\)~Regions found by the
enumeration (the completion did not finish).}
\label{tab:lti}
\end{table}

\begin{table}[!ht]
\centering\footnotesize\setlength{\tabcolsep}{3.5pt}
\begin{tabular}{l|cccc|c|cc|cc|cc}
\toprule
 & \multicolumn{4}{c|}{completed scenarios} & &
 \multicolumn{6}{c}{geometric mean of the speed-up (reference / CZ (It.\(\mid\)Q))}\\
 & CZ & CZ & & & paired & \multicolumn{2}{c|}{EO, \(T=1\)}
 & \multicolumn{2}{c|}{EO+CC, \(T=1\)} & \multicolumn{2}{c}{EO+CC, \(T=8\)}\\
\(n/m\) & EO & EO+CC & MPT3 & YALMIP & scenarios & MPT3 & YALMIP & MPT3 & YALMIP & MPT3 & YALMIP\\
\midrule
2/1 & 70 & 70 & 70 & 70 & 70 & 7.6 & 5.7 & 3.2 & 2.4 & 3.1 & 2.3 \\
3/1 & 70 & 70 & 70 & 56 & 56 & 6.5 & 4.8 & 2.6 & 1.9 & 2.7 & 2.0 \\
4/1 & 70 & 70 & 70 & 70 & 70 & 8.2 & 5.4 & 2.9 & 1.9 & 3.3 & 2.1 \\
4/2 & 68 & 61 & 58 & 59 & 58 & 5.8 & 4.7 & 2.8 & 2.3 & 3.7 & 3.0 \\
5/2 & 51 & 45 & 36 & 36 & 34 & 10.5 & 9.1 & 4.0 & 3.5 & 4.8 & 4.2 \\
\midrule
all & 329 & 316 & 304 & 291 & 288 & 7.4 & 5.5 & 3.0 & 2.3 & 3.3 & 2.5 \\
\bottomrule
\end{tabular}
\caption{All scenarios. Completed: out of \(70\) per class (\(5\) systems,
\(N=2,\dots,15\)) and \(350\) in all; CZ with \(T=1\). Speed-ups over the
paired scenarios, i.e.\ those completed by all methods, with the check.
Below one, the CZ method is slower.}
\label{tab:ltiagg}
\end{table}

\begin{table}[!ht]
\centering\small
\begin{tabular}{l|ccccc}
\toprule
\(n/m\) & \shortstack{scenarios\\checked} & \shortstack{with missed\\regions}
 & \shortstack{median missed [\%]\\(when \(>0\))} & \shortstack{max missed\\{[\%]}}
 & \shortstack{certified after\\completion}\\
\midrule
2/1 & 70 & 0 & -- & 0.0 & 70 \\
3/1 & 70 & 23 & 20.2 & 44.9 & 68 \\
4/1 & 70 & 47 & 6.9 & 49.0 & 60 \\
4/2 & 61 & 45 & 7.3 & 45.8 & 33 \\
5/2 & 45 & 39 & 2.9 & 28.5 & 20 \\
\midrule
all & 316 & 154 & 6.5 & 49.0 & 251 \\
\bottomrule
\end{tabular}
\caption{Completeness of the CZ enumeration on the random systems. Missed:
regions added by the completion, in \% of the completed partition.
Certified: no gap and no undecided point left after the completion.}
\label{tab:lticov}
\end{table}

\textbf{Completeness.} The answer to the question raised in Section~\ref{sec:platform} is
different for these systems than for the example of Section~\ref{sec:simulation}.
\begin{itemize}[leftmargin=1.5em]
\item The enumeration misses regions in \(154\) of the \(316\) checked
  scenarios (Table~\ref{tab:lticov}). For \(8\) of the \(25\) systems (the
  five with \(n/m=2/1\), \#21, \#24 and \#25) it finds every region at every
  horizon. The share of missed regions depends much more on the system than
  on the class: over the horizons, its maximum is below \(1\%\) for \(11\)
  systems, between \(2\) and \(12\%\) for \(6\), between \(15\) and \(35\%\)
  for \(5\), and \(45\)--\(49\%\) for three (\#22, \#42, \#61). The median
  over the scenarios with missed regions is \(6.5\%\).
\item This is consistent with the effect of degeneracy discussed in
  Section~\ref{sec:platform}: the missed regions are not reached from the regions found by a
  change of a single index, which is the only move of the top-down
  exploration.
\item After the completion, the partitions have the size of those of the
  references: the median difference to MPT3 and to YALMIP is zero. \(251\) of the \(316\) partitions are
  certified after the completion; in the others the \(10\) rounds leave a few
  gap points or undecided points.
\end{itemize}

\textbf{Computation time.}
\begin{itemize}[leftmargin=1.5em]
\item With the completion included, \texttt{CZ (Iterative | Quick)} is faster
  than MPT3 in all \(288\) paired scenarios and faster than YALMIP in \(283\)
  of them, also in the serial run: \(3.0\times\) and \(2.3\times\) in
  geometric mean with \(T=1\), \(3.3\times\) and \(2.5\times\) with \(T=8\)
  (Table~\ref{tab:ltiagg}). The margin is largest for the largest systems
  (\(n/m=5/2\): \(4.0\times\) and \(3.5\times\) serially).
\item The five serial cases where YALMIP is faster are at the largest
  horizons: \#42 at \(N=14,15\) (\(1.6\) and \(1.4\)\,s against \(1.4\) and
  \(1.2\)\,s), \#62 at \(N=14,15\) (\(141.6\) and \(173.3\)\,s against
  \(127.6\) and \(137.6\)\,s) and \#84 at \(N=14\) (\(652.1\) against
  \(632.8\)\,s; also at \(N=15\), \(872.7\) against \(766.9\)\,s, which MPT3
  did not reach). 
\item The time of the enumeration alone is \(7.4\times\) and \(5.5\times\)
  shorter than MPT3 and YALMIP serially, \(9.1\times\) and \(6.8\times\)
  with \(T=8\), and shorter in every paired scenario. These factors are an
  upper bound of the gain, since the enumeration alone may return an
  incomplete partition.
% \item The check is a large part of EO+CC: a median of \(57\%\) of the serial
%   time, and more when many regions are missing (\#61 at \(N=7\): \(140\)\,s
%   against \(22\)\,s of enumeration). It processes the regions one by one in
%   MATLAB, and the points that fall in gaps are located by a search over all
%   regions; it has not been optimized to the extent of the enumeration.
\item Within the same budget, the CZ method solves larger problems
  (Table~\ref{tab:ltiagg}): for \(n/m=4/2\) it completes \(68\) scenarios
  (\(61\) with the check) against \(58\) for MPT3 and \(59\) for YALMIP, for
  \(5/2\) models, it solves \(51\) (\(45\)) against \(36\) and \(36\). The largest scenario
  completed by all three methods is \#64 at \(N=15\), with \(51\,577\)
  regions: \(726\)\,s serially with the check (\(370\)\,s with \(T=8\)),
  against \(1\,234\)\,s for MPT3 and \(861\)\,s for YALMIP.
\end{itemize}

\textbf{Conclusion.} For random systems: \texttt{CZ (Iterative | Quick)} with the
completeness check is faster than MPT3 in every case and than YALMIP in all
but five single threaded cases at the largest horizons, by \(2\)--\(4\times\) in
geometric mean, and it reaches larger problems within the same time limits. The
random systems also show that the enumeration alone does not always find all
regions, so the completion step is needed for general systems; its cost is
the main target of future work on the implementation.

\FloatBarrier

\section{Conclusions and outlook}\label{sec:outlook}

The analysis shows that time spent by the fastest method is
split roughly evenly between the compiled kernels and MATLAB-level work
(memory allocation of small structures, cell-array management, interpreted
loops over the candidates). We expect further gains from future updates of the implementation
and from a compiled or natively multithreaded language. the example of Section~\ref{sec:simulation}, with the same
algorithm, the same partitions and no gap found by the a posteriori
completeness check, \texttt{CZ (Iterative | Quick)} is faster than both
reference implementations at all horizons considered, also when every method
runs on a single thread, and its independent child computations let it use
additional cores, which the reference implementations do not. The enumeration alone does not always find all
regions, and the completion step is needed. With this step included,
\texttt{CZ (Iterative | Quick)} remains faster than MPT3 in every case and
than YALMIP in all but five serial cases at the largest horizons
(\(3.0\times\) and \(2.3\times\) in geometric mean with one thread), and it
solves larger problems within the same time bounds.

\end{document}